\documentclass[psfig]{aa}
\usepackage{graphics,rotating,psfig,afterpage,multirow}
\begin{document}
\title{Crystalline silicate dust around
evolved stars\thanks{
Based on observations with ISO, an ESA project with instruments
funded by ESA Member States (especially the PI countries: France,
Germany, the Netherlands and the United Kingdom) and with the
participation of ISAS and NASA}
}
\subtitle{II. The crystalline silicate complexes}

\author{F.J. Molster\inst{1,2,\dagger}, L.B.F.M. Waters\inst{1,3}, A.G.G.M. Tielens\inst{4}}
\institute{Astronomical Institute `Anton Pannekoek', University of
Amsterdam, Kruislaan 403, NL-1098 SJ Amsterdam, the Netherlands
\and
School of Materials Science and Engineering, Georgia Tech, Atlanta, GA 30332-0245, USA
\and
Instituut voor Sterrenkunde, Katholieke Universiteit Leuven, Celestijnenlaan
200B, B-3001 Heverlee, Belgium
\and
SRON Laboratory for Space Research Groningen, P.O. Box 800,
NL-9700 AV Groningen, The Netherlands
}

\offprints{F.J. Molster: fmolster@so.estec.esa.nl\\
\mbox{ }$\, \dagger \!\!$ Present address: F.J. Molster, ESA/ESTEC, SCI-SO, Postbus 299, 2200 AG  Noordwijk, The Netherlands}

\date{received date; accepted date}

\authorrunning{F.J. Molster et al.}
\titlerunning{The crystalline silicate complexes}

\abstract{
This is the second paper in a series of three in which we present an
exhaustive inventory of the solid state emission bands
observed in a sample of 17 oxygen-rich dust shells
surrounding evolved stars. The data were taken with the Short and
Long Wavelength Spectrographs on board of the Infrared Space
Observatory (ISO) and cover the 2 to 200 $\mu$m wavelength range.
Apart from the broad 10 and 18 $\mu$m bands that
can be attributed to amorphous silicates, at least 49
narrow bands are found whose position and width indicate they
can be attributed to crystalline silicates. Most of these emission bands are
concentrated in well defined spectral regions (called complexes).
We define 7 of these complexes; the 10, 18, 23, 28, 33, 40 and 60
micron complex.
We derive average properties of the individual bands.
Almost all of these bands
were not known before ISO. Comparison with laboratory data suggests that
both olivines (Mg$_{2x}$Fe$_{(2-2x)}$SiO$_4$) and pyroxenes
(Mg$_x$Fe$_{(1-x)}$SiO$_3$) are present, with x close to 1, i.e.
the minerals are very Mg-rich and Fe-poor. This composition is
similar to that seen in disks surrounding young stars and in the solar
system comet Hale-Bopp. A significant fraction of the emission bands
cannot be identified with either olivines or pyroxenes. Possible other
materials that may be the carriers of these unidentified bands are
briefly discussed.
There is a natural division into objects
that show a disk-like geometry (strong crystalline silicate bands), and
objects whose dust shell is characteristic of an outflow (weak
crystalline silicate bands). In particular, stars with 
the 33.5~$\mu$m olivine band
stronger than about 25 percent over continuum are invariably disk sources.
Likewise, the 60 $\mu$m region is
dominated by crystalline silicates in the disk sources, while it is
dominated by crystalline H$_{2}$O ice in the outflow sources.
We show that the disk and outflow sources have 
significant differences in the shape of the emission bands. This difference 
must be related to the composition or grain
shapes of the dust particles. 
The incredible richness of the crystalline silicate spectra observed by
ISO allows detailed studies of the mineralogy of these dust shells, and
is the origin and history of the dust.}

\maketitle



\section{Introduction}

\label{sec:obser}

At the end of their life both low and high mass stars loose a
large fraction of their mass in the form of a dense stellar wind. When the
temperature in the outer regions of the atmosphere becomes low enough
solid material ({\em dust}) condenses out of the gas. 
Mass loss can eventually dominate stellar evolution. 
Since the dust may also play an important role in the mass loss process,
it is interesting to investigate the physical and chemical processes
responsible for dust formation.
One way to do this is to study the endproducts of this dust formation process. 
The composition of the dust which has condensed
provides valuable information on the conditions when the dust was formed and
thereafter.

The dust around stars can be observed at different wavelengths.
In the visible and near-infrared~(NIR)
one can look at the wavelength dependence of the absorption caused by
the dust. Another way to investigate the dust at these wavelengths is
by means of scattered light of the central source by small dust particles in
the circumstellar environment or with polarimetric
observations. In this paper we will study the emission of the dust in
the mid-infrared (MIR) and far-infrared~(FIR).

The dusty environments around evolved stars can be divided in carbon-rich
(C-rich) and oxygen-rich (O-rich) environments, depending on the
C/O ratio of the mass-losing stars. This division is the result of the
stability of the CO molecule, which is formed before the dust condenses.
If there is more carbon than oxygen (C/O $>1$) all
the oxygen will be trapped in CO, and the dust species will be carbonaceous,
e.g. SiC, Polycyclic Aromatic Hydrocarbons (PAHs) or amorphous carbon.
If the C/O ratio is smaller than one, i.e. there is more oxygen than carbon,
all the carbon is trapped in CO and O-rich dust will be formed,
e.g. simple oxides and silicates.

In 1995 the Infrared Space Observatory
(ISO; Kessler et al. 1996) was launched which opened the possibility
to study mass-losing stars at infrared wavelengths with unprecedented
wavelength coverage and spectral resolution.
Before the launch of ISO it was generally assumed that in
the dusty winds of O-rich evolved stars only amorphous silicates
were formed.
One of the remarkable discoveries of ISO was the detection of crystalline
silicates outside our own Solar System. Their spectral signature was not only
found in the spectra of young stars (Waelkens et al, 1996), but also in the
outflows of evolved stars (Waters et al, 1996). In the latter case, we even
found one example, IRAS09425-6040, where 75\% of the circumstellar dust 
consists of crystalline silicates (Molster et al. 2001).

In contrast to amorphous silicates, crystalline silicates provide a unique
opportunity to determine for the first time the chemical composition of the
dust particles. The relatively sharp features of the crystalline silicates
are very sensitive to compositional changes, this in contrast to the broad
and smooth amorphous silicate features.
Crystalline silicates will also help us to better understand the physical
and chemical conditions under which dust is formed. In particularly, they need
high temperatures and a slow cooling down to form.

Crystalline silicates are found in the outflows of evolved stars which
replenish the ISM, and also around young stars which form from the ISM
(e.g. Waters et al. 1996; Waelkens et al. 1996).
However, up to now no crystalline silicates have been found in the ISM.
The abundance in the ISM might be too low. However, this would make
it difficult to explain
the high abundance of crystalline silicates in a young star as HD100546
(Waelkens et al. 1996; Malfait et al. 1998). On the other hand, 
if the crystalline
silicates are destroyed in the ISM then the question arises how they are
formed around young stars. Possible solutions to this last problem are given
by Molster et al. (1999a; hereafter MYW).

The crystallization process is not well understood and 
in order to better constrain the origin of these grains, it is important to
accurately describe the observed properties in different environments.
After the first reports of the crystalline silicates, only
a few objects have been analysed in some detail, e.g. HD100546
(Malfait et al. 1998), CPD$-56^{\circ}8032$ (Cohen et al. 1999),
AFGL4106 (Molster et al. 1999b), IRAS09425-6040 (Molster et al. 2001).
This will be the first attempt to obtain an average spectrum of the
different crystalline silicate features which are found around evolved stars.

In the first paper of this series (Molster et al. submitted; hereafter paper I)
we have described the infrared spectra of 17 sources. Here, we will derive
mean spectra based on these sources and identify the different features.
In the third paper of this series (Molster et al. submitted; hereafter paper III)
we will apply a simple dust emission model to the spectra of our sample
and derive several (spectral) trends.

In Section~\ref{sec:obser} we will briefly summarise how we obtained these
mean spectra. For a more extensive description we refer to paper I.
The results will be shown in Section~\ref{sec:results},
where we will also order and define the different solid state bands and
complexes found. In Section~\ref{sec8:ident} we will identify most
of the features found in Section~\ref{sec:results} by comparison to
laboratory spectra of cosmic dust analogues.

\begin{table*}[b]
\caption[]{The characteristics of the different features. The mean $\lambda$ (FWHM)
is the average wavelength (Full Width Half Max) of the feature,
with the errors taken into account. $\lambda$ (FWHM) min and max are the minimum and maximum
value found in the sample. The N indicates not a member of one of the 7 complexes}
{\small
\begin{tabular}{|lrr|rrr|l|c|}
\hline
\multicolumn{3}{|c|}{$\lambda$}& \multicolumn{3}{|c|}{FWHM}& Identification & complex\\
mean & min & max & mean & min & max & & nr\\
\hline
\multicolumn{1}{|c}{8.3} &     8.20 &     8.40 &     .42 &      .41 &      .43 & & 10 \\
\multicolumn{1}{|r}{9.14} &     9.12 &     9.17 &     .30 &      .24 &      .68 & silica? & 10 \\
\multicolumn{1}{|r}{9.45} &     9.45 &     9.46 &     .19 &      .15 &      .25 & & 10 \\
\multicolumn{1}{|c}{9.8} &     9.77 &     9.84 &     .17 &      .14 &      .29 & forsterite + enstatite & 10 \\
  10.1 &     9.59 &    10.61 &    2.56 &     1.30 &     3.77 & amorphous silicate& 10 \\
  10.7 &    10.57 &    10.90 &     .28 &      .11 &      .66 & enstatite & 10 \\
  11.05 &    11.04 &    11.06 &     .05 &      .03 &      .11 & instrumental artifact & 10 \\
  11.4 &    11.33 &    11.50 &     .48 &      .38 &      .86 & forsterite, diopside? & 10 \\ 
\hline
  15.2 &    15.00 &    15.42 &     .26 &      .13 &      .73 &  enstatite & 18 \\
  15.9 &    15.69 &    16.06 &     .43 &      .24 &      .65 &  silica?? & 18 \\
  16.2 &    16.10 &    16.37 &     .16 &      .08 &      .62 &  forsterite & 18 \\
\multicolumn{1}{|l}{16.50}&    16.49 &    16.50 &     .11 &      .11 &      .11 &  PAH?& 18 \\
  16.9 &    16.73 &    17.07 &     .57 &      .37 &      .84 &  a blend of 16.7 and 17.0 & 18 \\
  17.5 &    16.79 &    18.46 &    2.10 &      .81 &     3.66 &  amorphous silicate      & 18 \\
  17.5 &    17.43 &    17.61 &     .18 &      .13 &      .36 &  enstatite & 18 \\
  18.0 &    17.90 &    18.16 &     .48 &      .28 &     1.24 &  enstatite + forsterite & 18 \\
  18.9 &    18.43 &    19.17 &     .62 &      .36 &     1.20 &  forsterite? & 18 \\
  19.5 &    19.36 &    19.75 &     .40 &      .14 &      .86 &  forsterite + enstatite & 18 \\
\hline
  22.4 &    22.26 &    22.51 &     .28 &      .16 &      .55 &   & 23 \\
  23.0 &    22.82 &    23.14 &     .48 &      .28 &      .72 &  enstatite & 23 \\
  23.7 &    23.45 &    23.81 &     .79 &      .54 &     1.29 &  forsterite & 23 \\
  23.89&    23.88 &    23.90 &     .18 &      .13 &      .25 &   & 23 \\
  24.5 &    24.16 &    24.65 &     .42 &      .16 &     1.04 &  enstatite + ? & 23 \\
  25.0 &    24.83 &    25.14 &     .32 &      .25 &      .53 &  diopside? & 23 \\
\hline
  26.8 &    26.71 &    26.93 &     .37 &      .21 &      .47 &   & 28 \\
  27.6 &    27.46 &    27.79 &     .49 &      .28 &     1.18 &  forsterite & 28 \\
  28.2 &    27.97 &    28.45 &     .42 &      .23 &      .90 &  enstatite   & 28 \\
  28.8 &    28.68 &    28.88 &     .24 &      .19 &      .42 &   & 28 \\
  29.6 &    29.37 &    29.90 &     .89 &      .58 &     1.99 &  2 features?, diopside? & 28 \\
  30.6 &    30.48 &    30.77 &     .32 &      .18 &      .81 &   & 28 \\
  31.2 &    31.12 &    31.27 &     .24 &      .21 &      .36 &  forsterite? & 28 \\
\hline
  32.2 &    32.06 &    32.51 &     .46 &      .24 &      .75 &  diopside? & 33 \\
  32.8 &    32.56 &    33.03 &     .60 &      .36 &     1.00 &   & 33 \\
  32.97&    32.96 &    32.99 &     .20 &      .11 &      .28 &  instrumental artifact & 33 \\
  33.6 &    33.45 &    33.71 &     .70 &      .52 &     1.15 &  forsterite & 33 \\
  34.1 &    33.93 &    34.36 &     .36 &      .17 &      .74 &  enstatite + diopside? & 33 \\
  34.9 &    34.67 &    35.35 &    1.36 &      .63 &     1.88 &  clino-enstatite? & 33 \\
  35.9 &    35.76 &    36.20 &     .53 &      .37 &      .88 &  ortho-enstatite? & 33 \\
  36.5 &    36.44 &    36.72 &     .39 &      .25 &      .97 &  forsterite + ? & 33 \\
\hline
  39.8 &    39.44 &    40.36 &     .74 &      .21 &     2.57 &  diopside? & 40 \\
  40.5 &    40.34 &    40.80 &     .93 &      .53 &     1.53 &  enstatite (a blend?) & 40 \\
  41.8 &    41.52 &    42.11 &     .72 &      .42 &     1.73 &  41 micron plateau & 40 \\
  43.0 &    42.55 &    43.07 &     .89 &      .51 &     1.59 &  cryst. H$_2$O-ice + clino-enst. & 40 \\
  43.8 &    43.30 &    44.05 &     .78 &      .41 &     3.01 &  ortho-enstatite & 40 \\
  44.7 &    44.39 &    45.13 &     .58 &      .42 &     1.16 &  clino-enstatite, diopside? & 40 \\
\hline
  52.9 &    51.40 &    56.55 &    3.11 &     1.75 &     5.91 &  cryst. H$_2$O-ice & 60 \\
  62.  &    61.24 &    65.56 &   11.80 &     4.67 &    15.29 &  cryst. H$_2$O-ice (62 micron) + & 60 \\
       &          &          &         &          &          &  enstatite?, diopside? (65 micron) & 60 \\
  69.0 &    68.79 &    69.15 &     .63 &      .46 &     1.04 &  forsterite & 60 \\
\hline
  13.5 &    13.40 &    13.60 &     .25 &      .17 &      .44 & responsivity  & N \\
  13.8 &    13.74 &    13.90 &     .20 &      .16 &      .23 & enstatite, responsivity?  & N \\
  14.2 &    14.15 &    14.28 &     .28 &      .18 &      .55 & enstatite, responsivity?  & N \\
  20.7 &    20.54 &    20.84 &     .31 &      .16 &      .84 & silica?, diopside? & N \\
  21.5 &    21.35 &    21.65 &     .35 &      .15 &      .79 &  & N \\
  26.1 &    25.91 &    26.29 &     .57 &      .22 &     1.10 & forsterite + silica? & N \\
  38.1 &    37.83 &    38.21 &     .57 &      .56 &      .61 &  & N \\
  47.7 &    47.55 &    47.83 &     .97 &      .78 &     1.85 & FeSi?, a silicate  & N \\
  48.6 &    48.34 &    49.07 &     .61 &      .44 &     1.37 & a silicate  & N \\
  90.9 &    89.59 &    91.12 &   14.44 &    12.72 &    17.63 &  & N \\
\hline
\end{tabular}}
\label{tab:complex}
\end{table*}

\begin{table*}
\caption[]{The presence of the features which are seen in at least
three different SWS spectra of our sample, + means emission, - is absorption,
o means no good data available, 2 implies that a feature split in 2
separate features in this source and underlined x's mean all present,
but measured as one feature. 
For MWC922 the blend of the 9.1 and 9.5 and for OH26.5+0.6
the blend of the 23.0 and 23.7 micron features are
in absorption, while all other blends are in emission. The order of the stars 
is in decreasing strength of the blend of the 33.6 micron feature. 
am = amorphous silicate, d= diopside; (o/c)e = (ortho/clino)enstatite; 
f = forsterite; i = instrumental artifact; s= silica; 
si = a crystalline silicate; w = crystalline water-ice. 
}
{\normalsize
\begin{tabular}{|l|c@{\ \ }c@{\ \ }c@{\ \ }c@{\ \ }
c@{\ \ }c@{\ \ }c@{\ \ }c@{\ \ }c@{\ \ }c@{\ \ }c@{\ \ }
c@{\ \ }c@{\ \ }c@{\ \ }c@{\ \ }c@{\ \ }c@{\ \ }c@{\ \ }
c@{\ \ }c@{\ \ }c@{\ \ }c@{\ \ }c@{\ \ }c@{\ \ }c@{\ \ }
c@{\ \ }|}
\hline
               & & & & & & & & & & & & & & & & & & & & & & & & & &  \\
               & & & & & & & & & & & & & & & & & & & & & & & & & &  \\
               & & & & & & & & & & & & & & & & & & & & & & & & & &  \\
Bandname &
{ \begin{rotate}{90} 8.3   \end{rotate}  }& 
{ \begin{rotate}{90} 9.1 s?  \end{rotate}  }& 
{ \begin{rotate}{90} 9.5   \end{rotate}  }& 
{ \begin{rotate}{90} 9.8 f,e  \end{rotate}  }& 
{ \begin{rotate}{90} 10 (am.) \end{rotate} }& 
{ \begin{rotate}{90} 10.7 e \end{rotate}  }& 
{ \begin{rotate}{90} 11.4 f,d? \end{rotate}  }& 
{ \begin{rotate}{90} 15.2 e \end{rotate}  }&     
{ \begin{rotate}{90} 15.9 s? \end{rotate}  }&     
{ \begin{rotate}{90} 16.2 f \end{rotate}  }&     
{ \begin{rotate}{90} 16.9  \end{rotate}  }&     
{ \begin{rotate}{90} 17.5 e \end{rotate}  }&     
{ \begin{rotate}{90} 18 (am.) \end{rotate} }& 
{ \begin{rotate}{90} 18.0 f,e \end{rotate}  }&     
{ \begin{rotate}{90} 18.9 f? \end{rotate}  }&     
{ \begin{rotate}{90} 19.5 f,e \end{rotate}  }&     
{ \begin{rotate}{90} 20.7 s?,d? \end{rotate}  }&     
{ \begin{rotate}{90} 21.5  \end{rotate}  }&     
{ \begin{rotate}{90} 22.4  \end{rotate}  }&     
{ \begin{rotate}{90} 23.0 e \end{rotate}  }&     
{ \begin{rotate}{90} 23.7 f \end{rotate}  }&     
{ \begin{rotate}{90} 23.89 \end{rotate}  }&     
{ \begin{rotate}{90} 24.5 e \end{rotate}  }& 
{ \begin{rotate}{90} 25.0 d? \end{rotate}  }& 
{ \begin{rotate}{90} 26.1 f,s? \end{rotate}  }& 
{ \begin{rotate}{90} 26.8  \end{rotate}  }\\    
 \hline
IRAS09425       & & & & & &?& &+&+&+&+& & &+&+&+&+&+&+&+&+& &+&+&+&+ \\
NGC6537         & & & & & & & & & & &+& & &+&+&+&+&+& &+&+& &+&+&+&+ \\
NGC6302         & & & & & & & & & & & & & &+&+&+&+&+&+&+&+& &+&+& &  \\
MWC922          & &\multicolumn{2}{@{}c@{\ \ }}{\underline{x~~~x}}&-& &-& &+& & &2&+&-&+&+&+&+&+& &+&+&+&+&+&+&+ \\
AC Her          &+&+&+&+&+&+&+&+&+&+&+&+&?&+&+&+&+&+&+&\multicolumn{2}{@{}c@{\ \ }}{\underline{x~~~x}}&+&+&+& &  \\
HD45677         &+&+&+&+&+&+&+&+&+& &+&+&+&+&+&+&+&+&+&+&+& &+&+&+&+ \\
89 Her          &+&+&+&+&+&+&+&+&+&+&+& &+&+&+&+& & & &+&+& &\multicolumn{2}{@{}c@{\ \ }}{\underline{x~~~x}}& &  \\
MWC300          & & & & &-& & & & & &+&+&-&+&+&+&+& &+&+&+& &+&+&+&  \\
VY 2-2          & & & & &+& & & & & & & & & & & &+&+& &\multicolumn{3}{@{}c@{\ \ }}{\underline{x~~~x~~~x}}&\multicolumn{2}{@{}c@{\ \ }}{\underline{x~~~x}}& &  \\
HD44179         & & & & & & & & &+&+&+& &?&+&+&+&+&+& &+&+& &+&+&+&  \\
HD161796        & & & & &+& & &+&+&+& & &+&+&+&+&+&+& &+&+& &+&+&+&  \\
OH26.5+0.6      & & & & &-& & &-&-&-&-&-&-&-&-&-&-& &-&\multicolumn{2}{@{}c@{\ \ }}{\underline{x~~~x}}&-&-&-&-&- \\
Roberts 22      & & & & & & & & &\multicolumn{2}{@{}c@{\ \ }}{\underline{x~~~x}}&?&?&+&?&?&?&+&+& &\multicolumn{5}{@{}c@{\ \ }}{\underline{x~~~x~~~x~~~x~~~x}}& &o \\
HD179821        & & & & &+& & & &\multicolumn{2}{@{}c@{\ \ }}{\underline{x~~~x}}&+& &+&\multicolumn{2}{@{}c@{\ \ }}{\underline{x~~~x}}&+&+& &+&+&+& &\multicolumn{2}{@{}c@{\ \ }}{\underline{x~~~x}}&+&  \\
AFGL4106        & & & & &+& & & &\multicolumn{2}{@{}c@{\ \ }}{\underline{x~~~x}}&+& &+&+&+&+&+&+&+&+&+& &+&+&+&  \\
NML Cyg         & & & & &-& & &+&+&+&+& &?&+&+& &+&+& &+&+& &+& & &o \\
IRC+10420       & &+&+& &+&?&+&+&+&+&+& &+&\multicolumn{3}{@{}c@{\ \ }}{\underline{x~~~x~~~x}}&+&+& &+&+& &+& &+&  \\
\hline
               & & & & & & & & & & & & & & & & & & & & & & & & & &  \\
               & & & & & & & & & & & & & & & & & & & & & & & & & &  \\
               & & & & & & & & & & & & & & & & & & & & & & & & & &  \\
Bandname &
{ \begin{rotate}{90} 27.6 f \end{rotate}  }& 
{ \begin{rotate}{90} 28.2 e \end{rotate}  }& 
{ \begin{rotate}{90} 28.8  \end{rotate}  }& 
{ \begin{rotate}{90} 29.6 d? \end{rotate}  }& 
{ \begin{rotate}{90} 30.6  \end{rotate}  }& 
{ \begin{rotate}{90} 31.2 f? \end{rotate}  }& 
{ \begin{rotate}{90} 32.2 d? \end{rotate}  }& 
{ \begin{rotate}{90} 32.8  \end{rotate}  }&     
{ \begin{rotate}{90} 32.97 i \end{rotate}  }&     
{ \begin{rotate}{90} 33.6  f \end{rotate}  }& 
{ \begin{rotate}{90} 34.1  e,d? \end{rotate}  }&     
{ \begin{rotate}{90} 34.9  ce? \end{rotate}  }&     
{ \begin{rotate}{90} 35.9  oe? \end{rotate}  }&     
{ \begin{rotate}{90} 36.5  f \end{rotate}  }& 
{ \begin{rotate}{90} 39.8  d? \end{rotate}  }& 
{ \begin{rotate}{90} 40.5  e \end{rotate}  }& 
{ \begin{rotate}{90} 41.8  \end{rotate}  }& 
{ \begin{rotate}{90} 43.0  w,ce \end{rotate}  }& 
{ \begin{rotate}{90} 43.8  oe \end{rotate}  }& 
{ \begin{rotate}{90} 44.7  ce,d?\end{rotate}  }& 
{ \begin{rotate}{90} 47.7  si \end{rotate}  }&    
{ \begin{rotate}{90} 48.6  si \end{rotate}  }&    
{ \begin{rotate}{90} 52    w \end{rotate}  }&    
{ \begin{rotate}{90} 62    w,d \end{rotate}  }&    
{ \begin{rotate}{90} 69.0  f \end{rotate}  }&    
{ \begin{rotate}{90} 91     \end{rotate}  }\\   
 \hline
IRAS09425       &+&+& &+&+& &+&+& &+&+&+&+&+&+&+&+&+&+&+& & & & & & \\
NGC6537         &\multicolumn{2}{@{\ \ }c@{\ \ }}{\underline{x~~~x}}&+&+&+& &+&+& &+&+&\multicolumn{2}{@{}c@{\ \ }}{\underline{x~~~x}}&+&+&+&+&+&+&+&+&+&+&+&+&+\\
NGC6302         &+&+& &+&+&+&+&+& &+&+&+&+&+&+&+&+&+&+&+&+&+&+&+&+&+\\
MWC922          &+&+&+&+&+&+&+&+& &+&\multicolumn{2}{@{}c@{\ \ }}{\underline{x~~~x}}&+&+& &+&+&+&+&+&\multicolumn{2}{@{}c@{\ \ }}{\underline{x~~~x}}&+&+&+& \\
AC Her          &+&+& &+&+&?&+&+& &\multicolumn{2}{@{}c@{\ \ }}{\underline{x~~~x}}& &\multicolumn{2}{@{}c@{\ \ }}{\underline{x~~~x}}&\multicolumn{2}{@{}c@{\ \ }}{\underline{x~~~x}}& & &+& & & & & & & \\
HD45677         &+&+& &+&+& &+&+& &+&\multicolumn{2}{@{}c@{\ \ }}{\underline{x~~~x}}&\multicolumn{2}{@{}c@{\ \ }}{\underline{x~~~x}}&+&+&+&+&+&+& & & & & & \\
89 Her          &+&+& &?& & & &+& &+&\multicolumn{2}{@{}c@{\ \ }}{\underline{x~~~x}}& & & & & & & & & & & & & & \\
MWC300          &\multicolumn{2}{@{\ \ }c@{\ \ }}{\underline{x~~~x}}&+&+&+&+& &+& &+&+&\multicolumn{3}{@{}c@{\ \ }}{\underline{x~~~x~~~x}}& &+& &\multicolumn{3}{@{}c@{\ \ }}{\underline{x~~~x~~~x}}& & & & & & \\
VY 2-2          & & & &+&+& & &+& &+&+&+&+&+& &+&\multicolumn{3}{@{}c@{\ \ }}{\underline{x~~~x~~~x}}& & & &+&+& &+\\
HD44179         &+&+& &+&+& &+&+& &+&+&+&+&+& &+& & &+& &\multicolumn{2}{@{}c@{\ \ }}{\underline{x~~~x}}&+&+&+& \\
HD161796        &+&+&?&+&+& &+&+& &+&+&+&\multicolumn{2}{@{}c@{\ \ }}{\underline{x~~~x}}&+&+&+&+&?& &\multicolumn{2}{@{}c@{\ \ }}{\underline{x~~~x}}&+&2&+& \\
OH26.5+0.6      & & & &+& & & &\multicolumn{4}{@{}c@{\ \ }}{\underline{x~~~x~~~x~~~x}}&+&+&+&+&+&+&+&+&+&\multicolumn{2}{@{}c@{\ \ }}{\underline{x~~~x}}&+&+&+& \\
Roberts 22      &+&+&+&o& & & &+& &+&+&+&\multicolumn{2}{@{}c@{\ \ }}{\underline{x~~~x}}&\multicolumn{2}{@{}c@{\ \ }}{\underline{x~~~x}}&+&+&+& & & & &+&?&+\\
HD179821        &+&+& &+& & &+&+& &+&+&+&+& &+&+&+&+&+&+&+&+&+&+&+& \\
AFGL4106        &+&+& &o&o& &+&+&+&+&+&+&+&+&+&+&+&+&+&+&\multicolumn{2}{@{}c@{\ \ }}{\underline{x~~~x}}& &+& & \\
NML Cyg         &o&o&o&o&+& &+&+&+&+&+&+&+& &\multicolumn{2}{@{}c@{\ \ }}{\underline{x~~~x}}&+&+&+&+&\multicolumn{2}{@{}c@{\ \ }}{\underline{x~~~x}}& &+&+& \\
IRC+10420       &+&+& &o& & & &+&+&+& &+&+& &?&+&+&+& &+&+&+&+&+& & \\
\hline
\end{tabular}}
\label{tab:overview2}
\end{table*}

\section{The derivation of the mean spectra.}

\subsection{The sample}

We have selected 17 evolved stars which all showed the presence
of crystalline silicates in their infrared spectra. These 17
sources covered a wide range in evolutionary status. The list
included 2 AGB stars (OH26.5+0.6 and IRAS 09425-6040), 5 post-AGB
stars (HD44179, 89 Her, AC Her, Roberts 22 and HD161796), 3
planetary nebula (NGC6537, NGC6302 and VY~2-2), three massive
stars (IRC+10420, AFGL4106 and NML~Cyg) and 4 objects with poorly
known evolutionary status (MWC922, MWC300, HD45677 and HD179821).
An extensive discussion of the individual stars and their spectra
can be found in Paper I (Molster et al. 2002a).

For all stars in this sample, spectra were taken with the Short Wavelength
Spectrometer (SWS; de Graauw et al, 1996) on board ISO and for most of them
we also got spectra taken by the Long Wavelength Spectrometer (LWS; Clegg et
al. 1996).

As we will show later in this paper, there is a clear difference in the spectra
of stars which (are expected to) have a disk and those who have not.
This difference is seen not only in the strength of the features (MYW),
but as we will demonstrate also in the shape and position of the features 
(see below and Molster et al. 2002b, hereafter Paper~III). 
Therefore, we have decided to divide our sample 
accordingly into 2 groups. 10 stars in our sample belong to the 
`disk' sources and 7 stars belong to the `non-disk' or `outflow' 
sources.

Although the spectral shape of the different complexes is
different for the disk and the outflow sources, the individual
bands are present in both type of classes (although in different
strength ratios!). The average FWHM (full width at half maximum)
and position of individual bands is therefore determined based on
the complete sample and no discrimination has been made between
disk and outflow sources.

\subsection{Mean complex spectra}

For all stars we defined a continuum, which we subtracted from the data to
obtain a better view of the dust features; see Paper I for more details on how
we derived the continuum. The resulting continuum subtracted spectra were used
to determine the mean complex spectra.

In the (continuum subtracted) spectra of the stars, we can
identify 7 different complexes in which multiple components can
be identified; some of which are severely blended. These
complexes are found near 10, 18, 23, 28, 33, 40 and 60
$\mu$m, and named after their position, resp. the 10, 18, 23, 28,
33, 40 and 60 micron complexes.

In order to get a reference spectrum for the different complexes
we have derived `mean' spectra for each complex. The mean complex
spectra were derived for both the outflow and the disk-sources
separately. In order to obtain
these mean complex spectra we have extracted each complex from
all the continuum subtracted spectra. The complexes of the stars
belonging to one group (disk or outflow sources) were added,
using a weighing factor proportional to the S/N of that part of
the spectrum, to create a `mean' spectrum (see Paper I for more details).

\afterpage{\clearpage}

\section{Results for the complexes}
\label{sec:results}

In this section we will systematically discuss the general trends in
the spectra of our program stars, by sub-dividing the sample into disk-sources
and outflow-sources, and by dividing the spectra into the different
complexes: the 10, 18, 23, 28, 33, 40 and 60 micron complex.
Because of the low dust temperature of most sources and
therefore low flux levels below 7 $\mu$m and the
absence of crystalline silicate features in this wavelength range, we
decided to limit our investigation to wavelengths longwards of 7~$\mu$m.

First, we will give an overview of the results that apply to the whole sample.
Then, we will discuss the different complexes individually, after which we will
identify many of the features.
In the discussion of the different complexes and features, we will refer
to feature names as xx~micron features and to wavelength positions as
yy~$\mu$m, e.g. the 30.6 micron feature of IRAS09425-6040 is located
at 30.48 $\mu$m (Paper I).

In Table~\ref{tab:complex}, we have grouped all features seen in at least
2 stars and indicated their wavelength and FWHM limits.
In principle, we assigned one feature to one spectral structure.
If in at least two stars a spectral structure was split into two different
components, we assigned two features to it. If in other stars
these two features could not be separated we assumed them to be heavily
blended and measured them together.
The one exception is the
16.9 micron feature. In MWC 922 it was found to be split into two
different components, while in all other stars it was blended.
We allowed the features to show a small shift in wavelength position.
The ranges in shifts are indicated in Table~\ref{tab:complex} by
$\lambda_{\rm min}$ and $\lambda_{\rm max}$.
This method leads to a minimum number of features to analyse all spectra.
Measured blends and uncertain features were not taken into account for the
values in Table~\ref{tab:complex}.

In the same table we also report to which mineral the feature could
be traced; This will be discussed extensively in 
Section~\ref{sec8:ident}. We decided not to discuss the C-rich features which 
are found in many of the disk sources.

The mean spectra of the different complexes are shown in
Fig.~\ref{fig:mean10} to Fig.~\ref{fig:mean60}.
In the next subsections we discuss these different complexes for all stars.
We point out that the individual disk source and outflow source spectra
show significant source to source variations in shape and
strength of the spectrum. Nevertheless, clear trends can be seen that are most
obvious in the mean spectra.

Table~\ref{tab:overview2} gives a schematic quick-look overview of what
features are found in what star. We indicate by using different symbols
the blends and the features which are uncertain. The order of the stars 
is in decreasing strength of the blend of the 33.6 micron feature.

\subsection{10 micron complex (7 -- 13 $\mu$m)}
\label{sec:10mu}

\begin{figure}[t]
\centerline{\psfig{figure=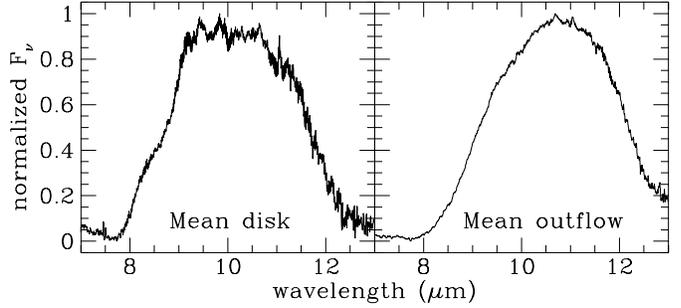,width=88mm}}
\caption[]{The normalised mean 10 micron complex spectrum of the disk and outflow sources.
The normalised mean spectrum for the disk sources is derived from AC Her, HD45677
and 89~Her, the normalised mean spectrum for the outflow sources is derived from all spectra
except OH26.5+0.6 and NML~Cyg.}
\label{fig:mean10}
\end{figure}

In Fig.~\ref{fig:mean10} the mean 10 micron complexes are shown.
The mean disk spectrum is only based on the 3 stars which show an
O-rich dust spectrum in this area, AC~Her, HD45677 and 89~Her.
All other disk sources have contributions of carbon-rich
material. The mean outflow spectrum is
based on the outflow sources which show the silicates in
emission, i.e. we have not used the absorption spectra of NML~Cyg
and OH26.5+0.6, because they have too strong absorption in this
wavelength range. The mean outflow spectrum is dominated by a
broad feature, that peaks at about 10.6~$\mu$m.

In the mean disk spectrum, several narrow features are found, at
8.3, 9.14, 9.45, 9.8, 10.7, 11.05 and 11.4~$\mu$m. We note
that it is likely that there is an underlying broad band 
in the 10 micron complex spectra of the disk
sources. Its shape and peak position seem somewhat different
from that seen in the outflow sources. However, we want to stress
that significant source to source variations are apparent in the
individual spectra (see Paper I).

\subsection{18 micron complex (15 -- 20 $\mu$m)}
\label{sec:18mu}

\begin{figure}[t]
\centerline{\psfig{figure=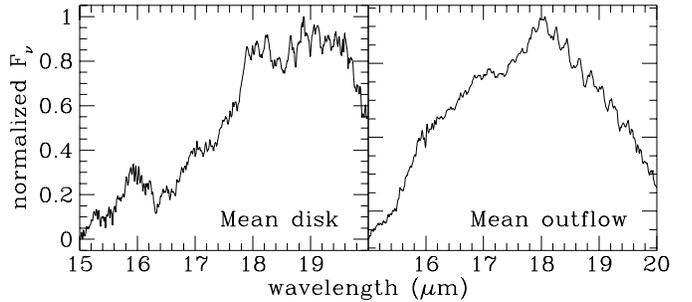,width=88mm}}
\caption[]{The normalised mean 18 micron complex spectrum of the disk and 
outflow sources. All stars, except NML~Cyg and OH26.5+0.6 for the
outflow sources, are used.} 
\label{fig:mean18}
\end{figure}

In Fig.~\ref{fig:mean18}, the mean 18 micron complexes are shown for the disk 
and outflow sources. 
OH26.5+0.6 has both the amorphous and crystalline silicates in
absorption in the 18 micron complex. Therefore, we have excluded this star 
from the mean spectrum. 
Also NML~Cyg shows the amorphous silicate in absorption in the 18 micron
complex, although the crystalline silicates seem already in emission we have
neglected this star for the mean outflow spectrum too. 
Excluding the two high flux, high S/N sources IRC+10420 
and AFGL4106, did not modify the average significantly. 
We note that some residual fringing is still visible in the average profile.

The 18 micron complex can be divided into narrow features at 15.2, 15.9
(with a shoulder at 16.2), 16.5, 16.9, 17.5, 18.0, 18.9 and 19.5 $\mu$m.
In both classes, the same narrow features are found. In all sources,
there are at least indications for the presence of amorphous silicates.
However, the outflow sources show a smoother complex due to the larger 
abundance of amorphous silicates (see also MYW).
The presence of the crystalline silicates (the narrow peaks)
are most prominent in the disk-sources.
Note the weakness or absence of the 19.5 micron
feature in the outflow sources.

The 18.1~micron feature might be a blend of 3 features. There is an indication 
for this in the spectrum of HD44179. The triple structure found in this 
spectrum is seen in all datasets available for this object. Also the
mean spectrum has a hint of these three features, indicating that they might be
weakly present in the individual spectra, close to the detection limit.

\subsection{23 micron complex (22. - 25.5 $\mu$m)}
\label{sec:23mu}

\begin{figure}[t]
\centerline{\psfig{figure=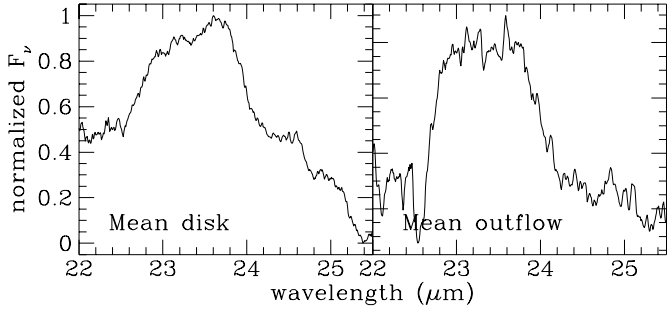,width=88mm}}
\caption[]{The normalised mean 23 micron complex spectrum of the disk and 
outflow sources. All sources are used except for OH26.5+0.6, which still has
all the features in absorption.}
\label{fig:mean23}
\end{figure}

The mean 23 micron disk and outflow complexes are shown in 
Fig.~\ref{fig:mean23}. The source OH26.5+0.6 still shows an absorption 
spectrum in the 23 micron complex for both the amorphous and crystalline 
silicates, therefore we have neglected this source for the calculation
of the mean outflow spectrum. 

The complex is much less influenced by the
presence of amorphous silicates and the crystalline silicate bands
are more prominent both for the disk and outflow sources.
The complex can be divided into 6 (blending) features at
22.4, 23.0, 23.7, 23.9, 24.5 and 25.0 $\mu$m.
These last two features form the 25 micron plateau.
The 23.9 micron feature is only seen in the
spectra of MWC 922, AC Her and OH26.5+0.6 (see Table~\ref{tab:overview2}
and Paper I). In the latter object, it is present in absorption.

The main difference between
the mean complexes of the disk and the outflow sources is
the relative strength of the 23.7
micron feature compared to the 23.0 micron feature, which is higher in the
disk sources than in the outflow sources. We note that 
variations exist from source to source within each group.

Another difference is that the spectra of disk sources show a strong 
24.5~micron feature in the 25 micron plateaus, while the spectra of
outflow sources do not. A third difference is the steep rise of the 23.0
micron feature, present in most outflow sources, but not in the
disk sources. Finally, we note that in almost all outflow sources
there is a sharp drop around 22.5 $\mu$m, while this is not found or at least
much less prominent in the disk sources.

\subsection{28 micron complex (26.5 -- 31.5 $\mu$m)}
\label{sec:28mu}

\begin{figure}[t]
\centerline{\psfig{figure=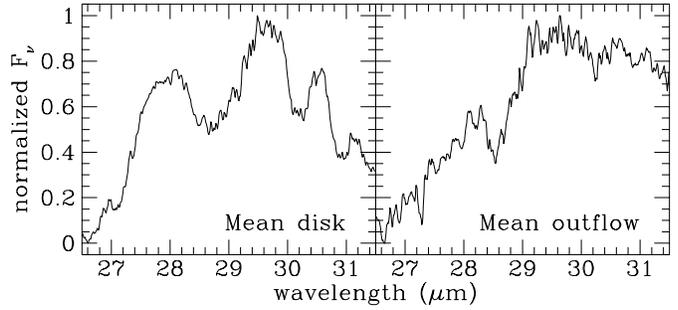,width=88mm}}
\caption[]{The normalised mean 28 micron complex spectrum of the disk and outflow sources.
All sources are used, except Roberts~22 for the mean disk spectrum and
AFGL 4106, NML~Cyg and IRC+10420 for the outflow sources, because of
limited wavelength coverage.}
\label{fig:mean28}
\end{figure}

In Fig.~\ref{fig:mean28}, we show the mean 28 micron complexes for
both the disk and outflow sources. For 4 sources, complete
wavelength coverage was not available. These stars (Roberts~22,
AFGL4106, NML~Cyg and IRC+10420) have therefore not been taken
into account in the mean disk and mean outflow spectra. This
especially influences the mean outflow spectrum, since, apart from
the 3 sources without a complete wavelength coverage, there is
one source which goes from absorption into emission around these
wavelengths and 2 sources are severely affected by noise.
Therefore the mean outflow spectrum in this wavelength region
should be treated with caution. This wavelength range is
dominated by SWS band 3E ($\approx 27.5-29.2~\mu$m), whose flux
and relative spectral response calibration are somewhat more
uncertain than for the other bands. Nevertheless, our large
sample permits us to draw conclusions about the features present
in this wavelength range.

This complex is built up from features at 26.8, 27.6, 28.2, 28.8,
29.6, 30.6 and 31.2~$\mu$m. The 29.6~$\mu$m feature might be a
blend itself, as indicated by the large variation in the peak
position (Table~\ref{tab:complex}). However, one should be careful
since there is an SWS band edge in the same region.

The 29.6 and 30.6~micron features seem related to each other. A
strong 30.6~micron features corresponds to a strong 29.6~micron
feature and vice versa.

\subsection{33 micron complex (31.5 -- 37.5 $\mu$m)}
\label{sec:33mu}

\begin{figure}[t]
\centerline{\psfig{figure=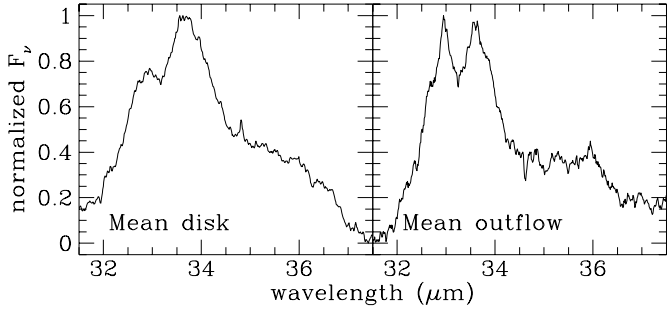,width=88mm}}
\caption[]{The normalised mean 33 micron complex spectrum of the disk and outflow sources.}
\label{fig:mean33}
\end{figure}

In Fig.~\ref{fig:mean33}, the 33 micron complexes are shown. The
shape of this complex bears similarities to the shape of the
23~micron complex. Waters et al. (1996) noted the presence of 3
main components: at 32.8~$\mu$m, at 33.7~$\mu$m and a plateau
extending up to 37~$\mu$m, hereafter called the 35~micron
plateau. However a closer look reveals much more structure. This
is most clearly seen in the spectrum of NML~Cyg (see Paper~I).
There is a weak feature at 32.2~$\mu$m. The 33.7~micron peak is
split into at least two features, peaking around 33.6 and
34.1~$\mu$m. Since these two features are seen in both the AOT01
and the AOT06 data of NML~Cyg and in the spectra of HD179821 and
AFGL4106 (see Paper~I) we are convinced that these structures are real.
The plateau shows some structure around 36 $\mu$m. It
seems that it can be split into three features, at 34.9, 35.9 and
36.5~$\mu$m. These features are in most cases severely blended and
just produce one broad feature (the 35 micron plateau).

In the infrared bright outflow sources, the
32.97~micron feature dominates, with IRC+10420 as the extreme case (see
Paper~I), while in the disk sources the 32.8~micron feature is relatively
strong, with AC Her as the extreme case (Paper~I). However, the 32.97~micron
feature coincides with a feature in the responsivity curve.
It is most prominently seen in the high flux sources and in our sample the
outflow sources are on average much brighter than our disk sources.
For comparison we also checked the bright C-rich source CRL2688, which
also shows the 32.97~$\mu$m feature. 
It is also seen in the spectrum
of other bright objects such as in Orion (Wright et al. 2000).
This suggests a flux-dependent shape of the band 4 responsivity curve, probably
related to band~4 memory effects. We conclude that the 32.97 micron feature
is instrumental.

We note that the
features seem sharper for the outflow sources than for the disk
sources. The (post-)Red Supergiants, which have the highest luminosity 
and mass loss rates, seem to have the sharpest peaks. Why
this is not the case for the other complexes is not clear yet.

The detailed shape of the plateau, especially at the short
wavelength side, is somewhat difficult to determine due to the
presence of the strong bands between 32.5 and 34.0~$\mu$m. The
red edge of the plateau falls at 36.5 -- 37~$\mu$m. This seems
correlated with the sharpness of the strong bands in this
complex. In the sources with the sharpest bands the plateau
extends to 36.5~$\mu$m, while in most other sources the plateau
extends to 37~$\mu$m.

The apparent difference between the disk and outflow sources in
the strength ratio of the 32.8 and 33.6~micron features is
totally due to the (post-)Red Supergiants, since the low mass
sources show about the same strength ratio as the disk sources.
It might be a good discriminator between (post-)AGB stars and
(post-)Red Supergiants.

\subsection{40 micron complex (38 -- 45.5 $\mu$m)}
\label{sec:40mu}

\begin{figure}[t]
\centerline{\psfig{figure=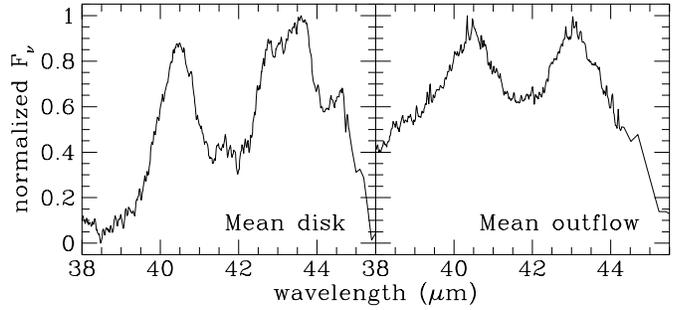,width=88mm}}
\caption[]{The normalised mean 40 micron complex spectrum of the disk and outflow sources.}
\label{fig:mean40}
\end{figure}

The mean 40 micron complexes (shown in Fig.~\ref{fig:mean40})
have 3 prominent features, one at 40.5~$\mu$m and the often
blended features at 43.0 and 43.8~$\mu$m. In addition to these
main features, there is a `plateau' visible from about 41 to
42.3~$\mu$m, hereafter called the 41~micron plateau, which is
probably a blend of features, which could not be separated.
Furthermore, at the short wavelength side of the 40.5~$\mu$m
feature, a shoulder is often seen around 39.7~$\mu$m and a weak feature is
present at 44.7~$\mu$m at the
long wavelength side of the 43.8~micron feature .

The main difference between the outflow- and the disk-sources
is seen in the strength of the 43.0 and 43.8~micron features. In
the outflow sources, the 43.0 micron feature is often stronger and
at least equally strong to the 43.8~micron feature, whereas in
the disk-sources the 43.8~micron feature often dominates the
43.0~micron feature. 

Another difference between the disk and outflow sources is the blue edge
of the 40.5~micron feature. In the outflow sources it has a more
gentle slope and it meets the continuum at wavelengths shortwards of
39.0~$\mu$m, while in the disk sources the slope is much steeper and ends
usually at wavelengths longwards of 39.0~$\mu$m.

\subsection{60 micron complex (50 -- 72 $\mu$m)}
\label{sec:60mu}

\begin{figure}[t]
\centerline{\psfig{figure=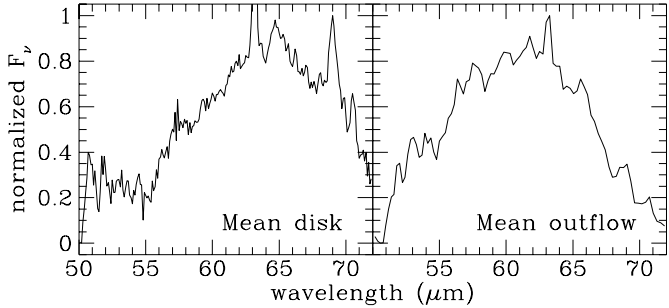,width=88mm}}
\caption[]{The normalised mean 60 micron complex spectrum of the disk and
outflow sources. The normalised mean spectrum for the disk sources is derived
from NGC6537, NGC6302, MWC922, HD44179 and Roberts 22, the normalised mean
spectrum for the outflow sources is derived from all outflow source spectra.}
\label{fig:mean60}
\end{figure}

For this part of the spectrum, not all sources were thoroughly reduced.
Therefore, we will only mention the global trends and not go into details.
The 60~micron complex (shown in Fig.~\ref{fig:mean60}) consists of a broad band
peaking at 60 -- 65~$\mu$m with a shoulder at 52~micron. In all disk sources,
except for Roberts~22 (see Paper~I), we find a prominent and sharp 69~micron
feature, while this feature is much less prominent in the outflow sources. 
It is interesting to note that Roberts~22 also shows similarities to outflow 
sources in other respects, i.e. low crystalline over amorphous ratio (see MYW).

There are indications of a feature at 57~$\mu$m, but this has not been measured
separately. The strong, narrow peak seen at 63.18~$\mu$m is due to the
fine-structure transition of [O I].

Disk and outflow sources differ in the peak position of the 60 -- 65~micron
band. In the outflow sources, it generally peaks around 62~$\mu$m, while in the
disk-sources it peaks at about 65~micron (with Roberts~22 as the exception).
The shift to longer wavelengths seems related to a decreasing 52~micron
shoulder, and increasing strength of the 29.6 and 30.6~micron features.

\subsection{Remaining features}

\begin{figure}[t]
\centerline{\psfig{figure=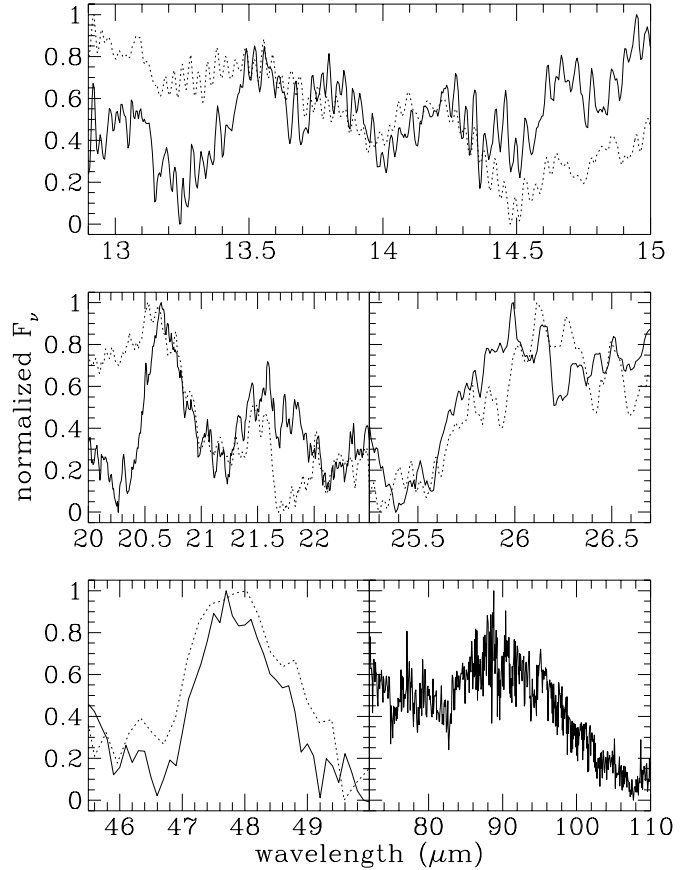,width=88mm}}
\caption[]{An overview of the remaining features, the dashed line
is the normalised mean spectrum for disk sources and the dotted line for
outflow sources.
There is no 90 micron feature detected in the
outflow sources.}
\label{fig:remain}
\end{figure}

In Fig.~\ref{fig:remain}, we show the mean spectra of the other features which
were not attributed to one of the complexes. Some of these features tend to
appear in most stars (C-rich and O-rich) and always in emission like the 
ones at 11.05, 13.6 (with a shoulder at 13.8) and 14.2 micron and are 
therefore believed to be (partly) instrumental. The strength of these features 
differs from source to source. However, it should also be noted that in some 
C-rich stars some of these features are seen up to 50\% times the continuum
level, which is far more than expected for an instrumental artifact. 
In these cases the 
features are attributed to PAHs (Tielens et al. 1999). For most of our sources 
however, we can exclude this possibility, since there is absolutely no 
evidence for PAHs at shorter wavelengths, which should be prominent if PAHs 
are present.

Real features are present at 20.7, 21.5, 26.1, 47.7 (with a shoulder
at 48.7) and 91 $\mu$m.

The objects HD45677 and OH26.5+0.6 both show a feature around
38~$\mu$m (see Paper I). The width of these two features is quite
similar, but the peak positions differ about 0.4~$\mu$m. Because of the
wavelength difference and because it is not seen in any of the
other stars in our sample, makes us doubt its reality.

\begin{figure*}[t]
\centerline{\psfig{figure=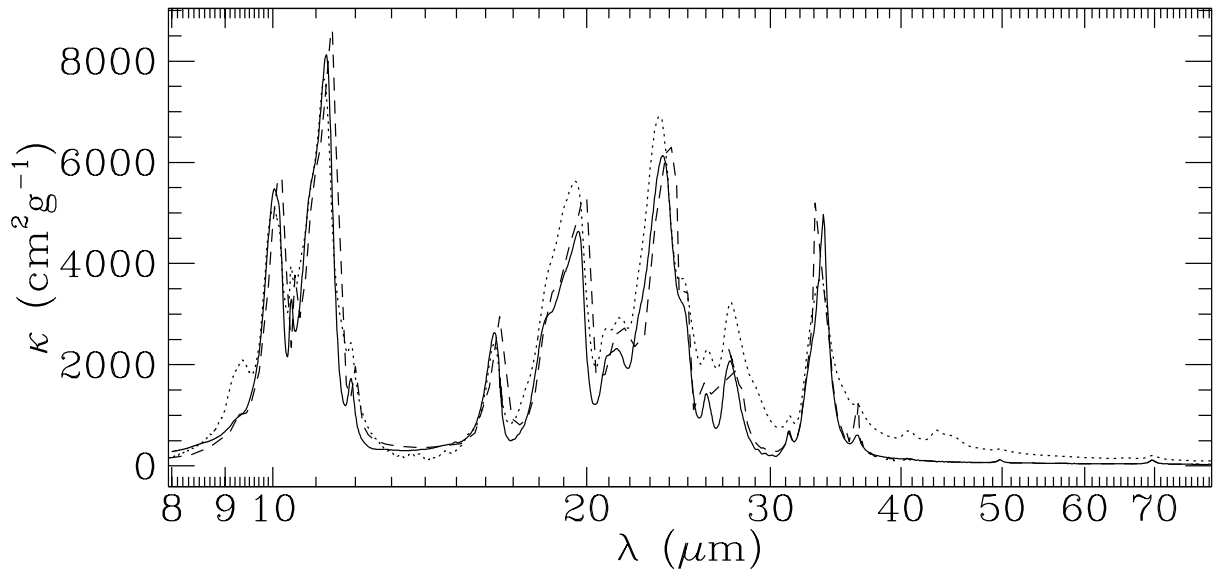,width=160mm,angle=270}}
\caption[]{The mass absorption coefficients of forsterite derived
from different laboratory measurements by, Koike et al. (1999; solid line),
J\"{a}ger et al. (1998; dotted line), Koike et al. (1993; dashed line).
The J\"{a}ger et al. data has been multiplied by a factor 2 to match the
other measurements}
\label{fig:forst}
\end{figure*}

\section{Identification}
\label{sec8:ident}

In this section, we will first discuss some of the laboratory
measurements of crystalline silicates that may be present in
circumstellar shells. Waters et al. (1996) identified most of
these features with olivines (Mg$_{2x}$Fe$_{(2-2x)}$SiO$_4$) and
pyroxenes (Mg$_{x}$Fe$_{1-x}$SiO$_3$), with $1\ge x \ge 0$. It
turned out that the best match with the observed features was
found for $x=1$ (see also Section~\ref{sec:ident60}); i.e. the
minerals forsterite (Mg$_2$SiO$_4$) and enstatite (MgSiO$_3$). We will
show the similarities and the differences between different
laboratory data sets of these materials. The origin of these
differences will be discussed. Finally, we will discuss the
identification for the different features per complex.

\subsection{Laboratory measurements of crystalline silicates}
\label{sec:lab}

\begin{figure*}[t]
\centerline{\psfig{figure=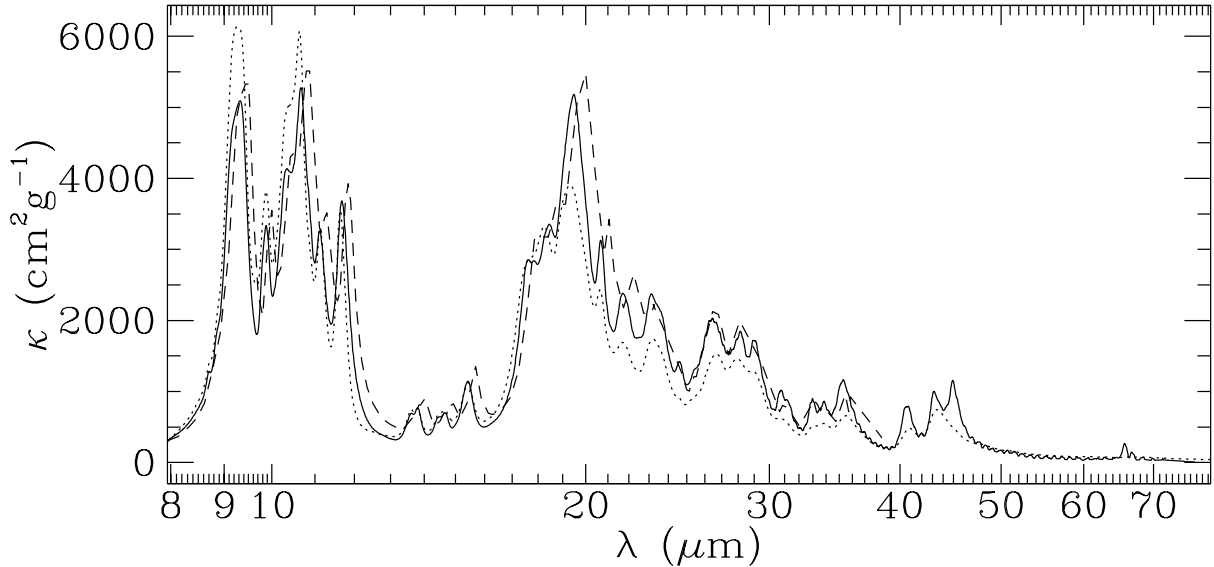,width=160mm,angle=270}}
\caption[]{The mass absorption coefficient of clino-enstatite derived from
different laboratory measurements by, Koike et al. (1999; solid line),
J\"{a}ger et al. (1998; dotted line), Koike \& Shibai (1998; dashed line).
This last data set only extends to 39~$\mu$m.}
\label{fig:c_enst}
\end{figure*}

\begin{figure*}[t]
\centerline{\psfig{figure=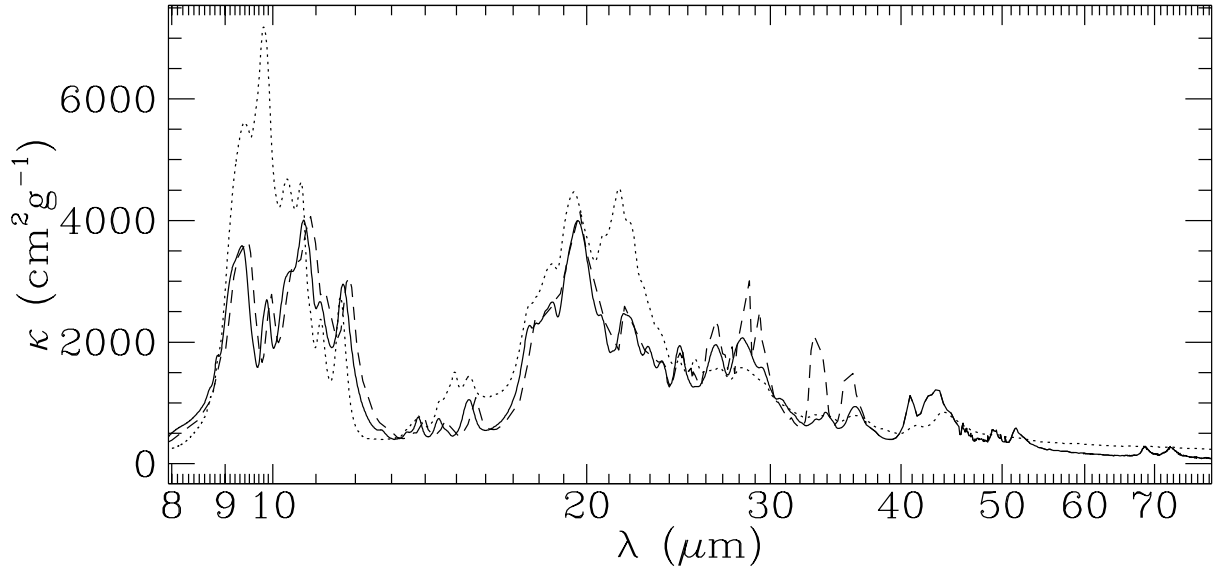,width=160mm,angle=270}}
\caption[]{The mass absorption coefficient of ortho-enstatite
derived from different laboratory measurements by,
Koike et al. (1999; solid line),
J\"{a}ger et al. (1998; dotted line), Koike \& Shibai (1998; dashed line).
The J\"ager et al.(1998) data is contaminated with talc at 9.8, 14.9 and 21.5
and 25.4 $\mu$m.}
\label{fig:o_enst}
\end{figure*}

In order to understand and identify the different features, which
were found in the infrared spectra of the dust around evolved
stars, good laboratory spectra are necessary. Most laboratory
spectra do agree on a global scale, but detailed comparison often
reveals quite some differences. In
Fig.~\ref{fig:forst},~\ref{fig:c_enst} and~\ref{fig:o_enst}, we
show different laboratory results for materials with the same
bulk composition. It is
clear that, although the average trends are very similar, there
are significant differences in the details; e.g., strength ratios.

The general trend in the 3 data sets of forsterite (taken from
J\"{a}ger et al. 1998, Koike et al. 1993 and Koike et al. 1999)
in Fig.~\ref{fig:forst} is the same. Forsterite has two strong
features in the 10~micron region, two prominent features in the
18 micron region, a strong band in the 23~micron complex and a
strong band in the 33~micron complex. Furthermore, there is a
weaker contribution in the 28~micron complex. A more detailed
examination reveals quite some differences between the laboratory
data sets. Peak positions are not identical, this might be a
grain shape effect, since the position (and shape) of the
forsterite features are grain shape dependent. In particular, the difference in
the peak position of the 23.7~micron feature for spherical
particles and for a continuous distribution of ellipsoids is
almost 1~$\mu$m. Therefore small differences between the
laboratory spectra and the ISO spectra can be expected and do not
lead automatically to a rejection of the proposed identifications. 
The J\"{a}ger et al. data set of forsterite shows extra features 
at 9.3, 40.7 and 
43.3~$\mu$m. These are probably due to enstatite inclusions. The measured 
absolute and relative strength of the features also differ. Note that we
rescaled the J\"{a}ger et al. data by a factor 2, in order to match the
other studies. At the moment, the origin of this difference is
unclear, but introduces substantial uncertainties in the derived relative
abundances (paper~III).

For enstatite there are two crystallographic configurations, the
rhombic structure (ortho-enstatite) and the monoclinic structure
(clino-enstatite). On Earth, ortho-enstatite is most common,
but enstatite found in meteorites and produced via melting in
laboratories is often of the clino-enstatite type. The structure
of the enstatite grains is related to its thermal history, which
makes them in principle a suitable candidate to explore the dust
conditions when these grains were formed. Below 40~$\mu$m, there
is not much difference between the spectra of clino- and
ortho-enstatite. The main differences are found in the 40 and
60~micron complexes.

In Fig.~\ref{fig:c_enst}, we plot the mass absorption coefficients
of clino-enstatite derived from three different laboratory
samples (J\"{a}ger et al. 1998; Koike \& Shibai 1998 and Koike et
al. 1999). Again, the general trend is very similar: a
lot of structure in the 10~micron complex, at least 7 different
features can be found here, a strong feature peaking slightly
longwards of 19~$\mu$m and decreasing to longer wavelengths with
a lot of substructure. Enstatite has several modes in the
40~micron complex, where forsterite is relatively smooth. In this
wavelength region, it also becomes possible to distinguish between
clino- and ortho-enstatite, which is very difficult below
40~$\mu$m, and this spectral region provides a prime probe for
enstatite particles. Unfortunately, the data sets do not
agree in this wavelength region. The sample preparation of the
J\"{a}ger 1998 sample is likely responsible for the absence of
the 44.7~micron feature in their data (J\"{a}ger, private
communication). A careful examination of the ISO spectra with the
laboratory data convinced us that the Koike 1999 data set (Koike
et al. 1999) best represents the features found in our ISO
spectra.

The spectrum of ortho-enstatite (Fig.~\ref{fig:o_enst}; data from
J\"{a}ger et al. 1998, Koike \& Shibai 1998 and Koike et al.
1999) is a clear example why the purity of the laboratory samples
always should be checked. The data from J\"{a}ger et al. (1998)
is contaminated by talc, producing various spurious
features. Up to 40~$\mu$m, the ortho-enstatite is very similar to
the clino- enstatite. The feature at 33.0~$\mu$m in the Koike
1993 data set (Koike et al. 1993) could not be reproduced by
other measurements and is therefore not believed to be real.
Therefore, the identification of ortho-enstatite to the 32.8~micron feature
by Waters et al. (1996) is reconsidered.

\begin{figure*}[t]
\centerline{\psfig{figure=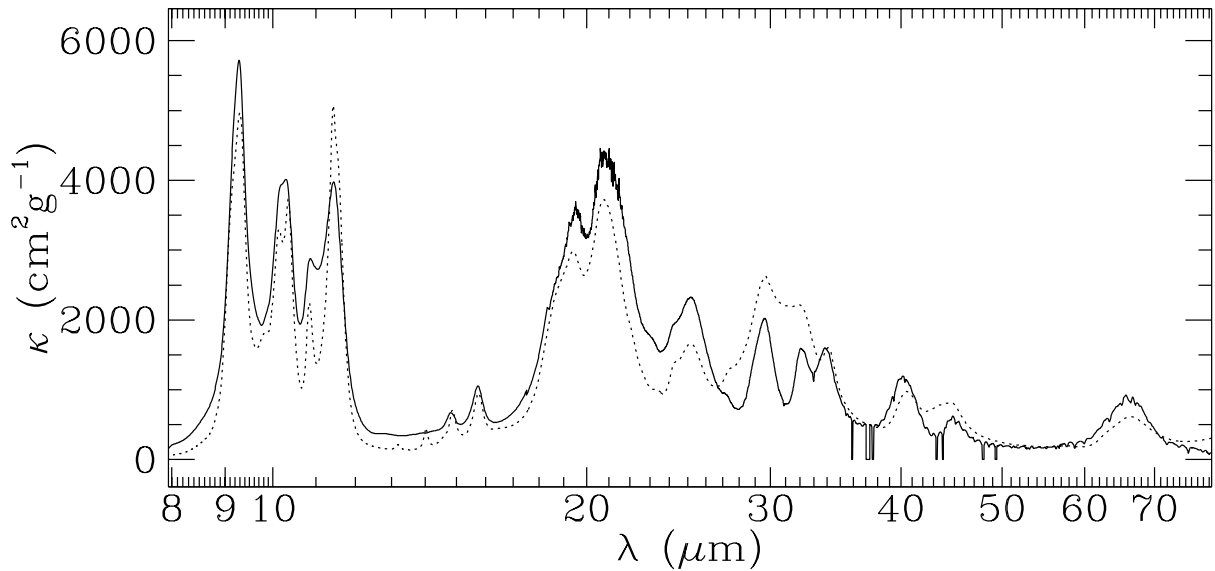,width=160mm,angle=270}}
\caption[]{The mass absorption coefficient of diopside
derived from different laboratory measurements by,
Koike et al. (1999; solid line) and J\"{a}ger (private communication).}
\label{fig:diop}
\end{figure*}

Diopside (MgCaSi$_2$O$_6$) is another silicate with possible
astrophysical relevance. This material shows peaks at 19.5,
20.6, 25.1, 29.6, 32.1, 33.9, 40.1, 44.8 and 65.7~$\mu$m. In
Fig.~\ref{fig:diop}, we show two laboratory measurements of this
material (Data from Koike et al. 1999 and J\"{a}ger private
communication). Again differences are found between these two
data sets, especially around 30~$\mu$m. 

These four examples show that one should be careful when applying a data set
from the literature, especially if it has not been checked on chemical
homogeneity, particle size and shape distribution. The homogeneity is not only
important for the natural samples, which often contain inclusions of other
materials (see e.g. J\"{a}ger et al (1998)), but synthetic samples are also 
not always homogeneous. 

It is important to realize that differences in the formation history of the 
minerals may lead to different structures and IR spectral characteristics.
Enstatite crystals which are formed from the vapour phase
in a low density hot gas environment often produce structures in which one of
the crystallographic axes is severely depressed (Bradley et al. 1983 and
references therein). This is thought to be the situation in the outflows of 
mass losing stars. Therefore the absence or weakness of one of the laboratory 
bands in the stellar spectrum might be related to this phenomenon.

In the following subsection, we will use the Koike 1999 data set for the
identifications, since that is thought to be the best and purest.
However, one should realize that small wavelength shifts (2\%) are found 
between the different laboratory samples, all of them claimed to be pure 
forsterite. Small defects in the ordering of the atoms might cause these 
differences. It is likely that similar differences are also found for the 
grains in space. One should also realize that the laboratory data sets were 
measured at room temperature, while most dust grains are much colder. A lower 
dust temperature often shifts the wavelengths of the peaks to shorter 
wavelengths and strengthen the features (see e.g. Henning \& Mutschke 1997,
Bowey et al. 2000).

\subsection{10 micron complex (7 -- 13 $\mu$m)}
\label{sec:ident10}

\begin{figure}[t]
\centerline{\psfig{figure=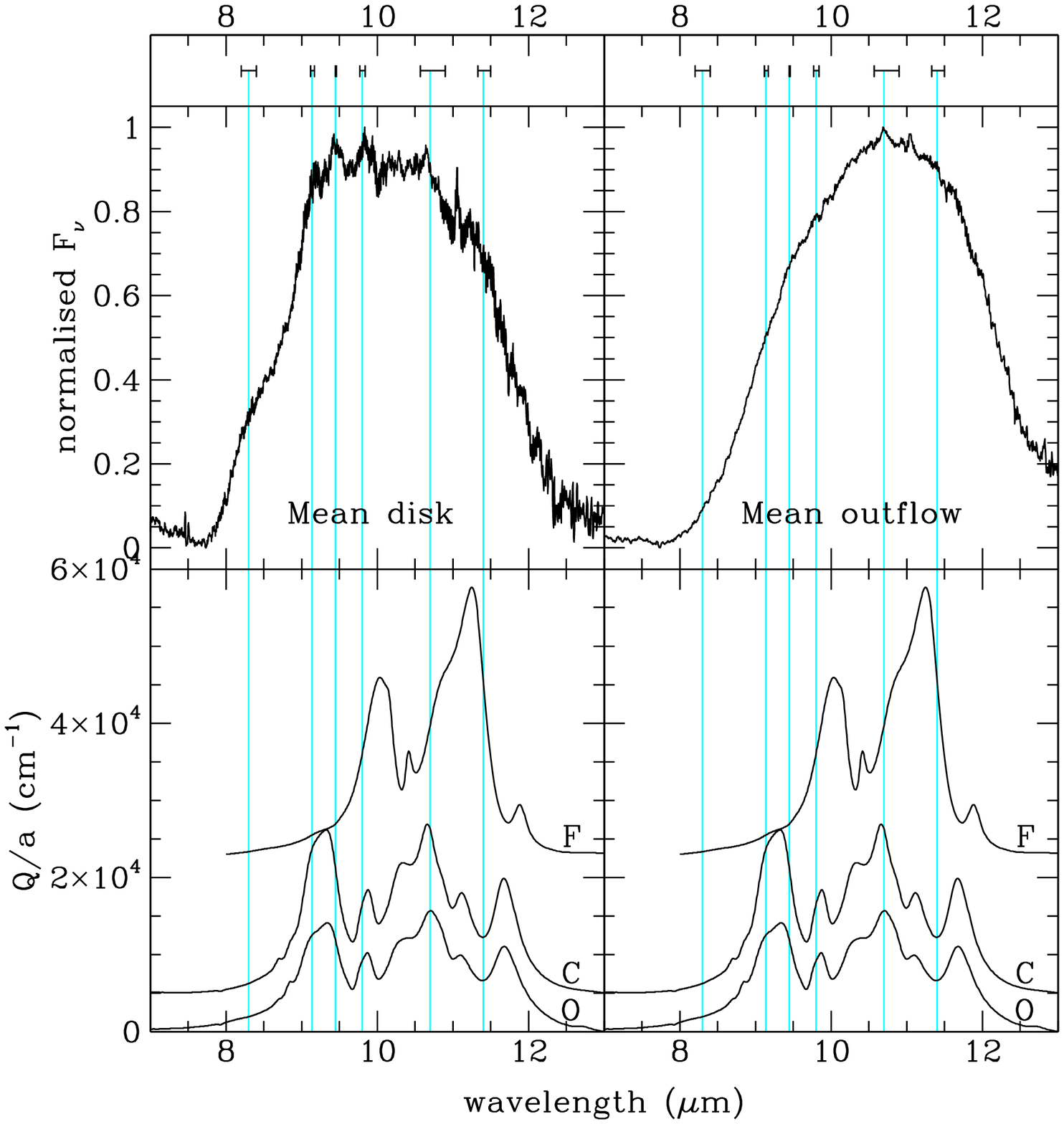,width=88mm,angle=0}}
\caption[]{The normalised mean 10 micron complex spectra together
with the
forsterite (F; +2.3E4) and clino-enstatite (C; +5E3) and ortho-enstatite (O) mass absorption coefficient
spectra in this wavelength range. The forsterite and clino enstatite
spectra are shifted. The gray lines indicate the features found in our
spectra, with their spread in peak position observed in the ISO data indicated above the plot.
The amorphous silicate peak has not been indicated.}
\label{fig:ident10}
\end{figure}

\begin{figure}[t]
\centerline{\psfig{figure=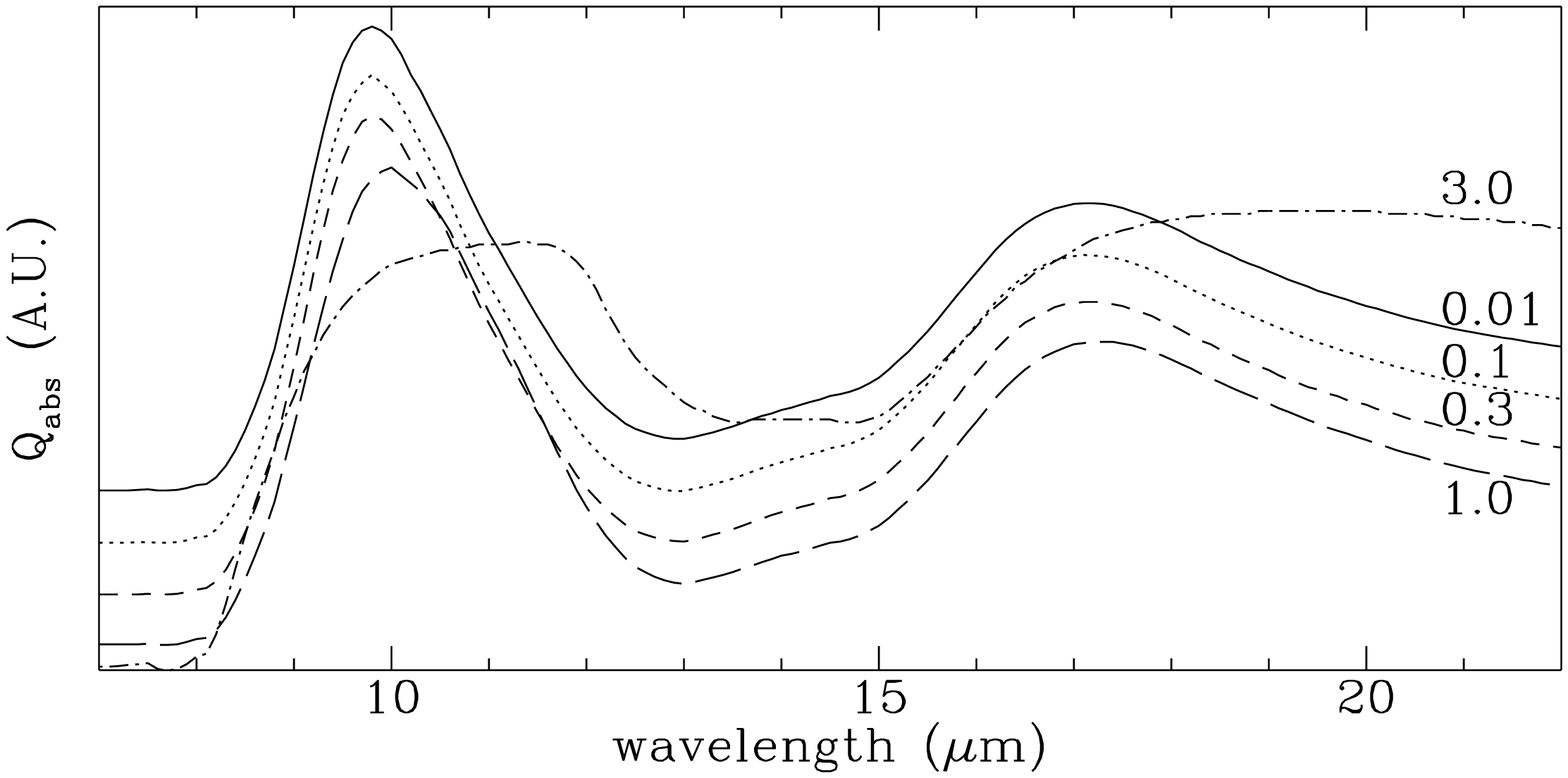,width=88mm,angle=270}}
\caption[]{The Q$_{\rm abs}$ for spherical olivine grains
(MgFeSiO$_4$; Dorschner et al. 1995) for different grain sizes:
0.01 $\mu$m (solid line), 0.1$\mu$m (dotted line),
0.3$\mu$m (short dashed line), 1.0$\mu$m (long dashed line), 3.0$\mu$m
(dashed dotted line).
The curves were scaled to get an equal strength around 10 $\mu$m and
then offset from each other.}
\label{fig:shift}
\end{figure}

In Fig.~\ref{fig:ident10}, we show the mean disk and outflow
source spectra together with the Koike 1999 data sets. We
subtracted a continuum from these data sets, in the same way as
we did for the infrared spectra of our sources, in order to get a
better comparison between the laboratory and ISO data. This
continuum subtracted data set is also used in the other
complexes. Note the similarities between the clino and
ortho-enstatite samples.

The broad feature in the 10~micron complex is thought to
originate from amorphous silicates. This feature peaks at a
relatively long wavelength (for both the average disk and outflow 
10~micron complex)
compared to the interstellar medium
amorphous silicate absorption feature (which is at 9.7~$\mu$m).
This can be an indication for the presence of large grains in the
dust around evolved stars 
(see e.g. Molster et al. 1999b). In Fig.~\ref{fig:shift}
we show this effect for spherical olivine grains (MgFeSiO$_4$;
Dorschner et al. 1995). Note the shift in peak position and the
broadening with increasing grain size. Besides 
grain size, the peak position and width of a feature can also be influenced 
by compositional changes
(see e.g. Tielens 1990; Dorschner et al. 1995; Demyk et al. 1999).
The redder peak position of the outflow sources is therefore
probably not simply due to the presence of larger grains. 
NB, it is the disk sources which do show evidence for 
very large grains (see e.g. Molster et al. 1999a). 

In the mean disk spectrum, several sharp features are present
which might be identified with crystalline silicates. The
forsterite feature, found in the laboratory data at 11.3~$\mu$m,
is slightly red-shifted, while the other strong forsterite
feature, at 10.0~$\mu$m in the laboratory data, is blue-shifted.
These shifts are within the differences
seen between the different laboratory data sets. According to the
laboratory data enstatite would peak between and around the
forsterite peaks, causing a severe blend of the features which
hampers identification of the separate features. The strong band
at 9.3~$\mu$m in the enstatite laboratory data seems split into
two components in the stellar spectrum. A hint of this splitting
is already seen in the laboratory data. The 10.7~micron band in
the ISO data coincides nicely with one of the strongest enstatite
bands. In the mean disk spectrum, the 11.4~micron band is likely a blend
of enstatite (at 11.7~$\mu$m) and the strong forsterite feature (at
11.3~$\mu$m). This blend might also explain
the (apparent) red-shift of the forsterite band in our ISO-data.

\subsection{18 micron complex (15 -- 20 $\mu$m)}
\label{sec:ident18}

\begin{figure}[t]
\centerline{\psfig{figure=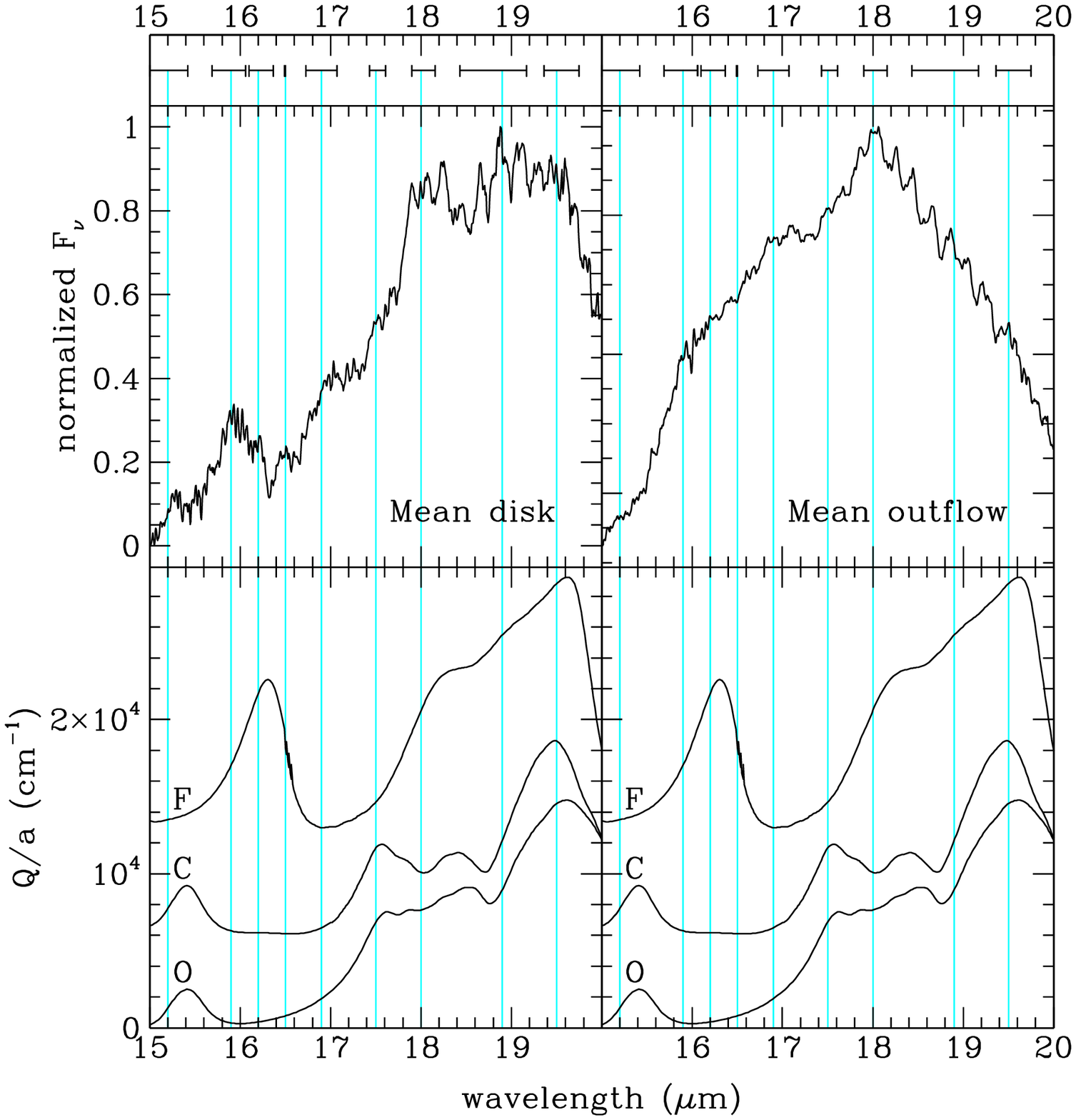,width=88mm,angle=0}}
\caption[]{The normalised mean 18 micron complex spectra together with the
forsterite (F; +1.3E4) and clino-enstatite (C; +6E3) and ortho-enstatite (O) mass absorption coefficient
spectra in this wavelength range. The forsterite and clino-enstatite
spectra are shifted. The gray lines indicate the features found in our
spectra, with their spread in peak position observed in the ISO data indicated above the plot.
The amorphous silicate peak has not been indicated.}
\label{fig:ident18}
\end{figure}

The mean spectra of the 18 micron complex are shown in Fig.~\ref{fig:ident18}
together with the laboratory spectra of forsterite and both types of enstatite.
Note that the difference between clino and ortho-enstatite is only marginal, 
implying that these two forms cannot be distinguished in this wavelength 
region.

Forsterite shows features around 16.3, 18.2 and 19.6 and a
shoulder at 19.0~$\mu$m (J\"{a}ger et al. 1998; Koike et al.
1999;), while enstatite shows features around 15.4, 17.6, 18.5
and 19.5~$\mu$m (J\"{a}ger et al. 1998; Koike et al. 1999). The
laboratory data of these silicates are only slightly
shifted from the peak positions found in our ISO-data. Because
all expected peaks are found, not only in this wavelength range
but also in the other complexes, the identification of enstatite
and forsterite is unambiguous. However, the peak strength of the
features is not as expected. According to the laboratory data the
peak at 19.5~$\mu$m should have been by far the strongest feature
in this complex. This is not observed and cannot be explained by
temperature effects. Since there are also unidentified features
at 16.0 and 16.9~$\mu$m it is likely that more dust components
are present. These dust species might also contribute at other
wavelengths in this complex and therefore change the apparent
strength ratios of the different features. In particular, the 18.1 and
18.9~micron features likely have a component underlying the
forsterite and enstatite features.

The 16.5 micron feature is only seen in two stars (NGC6537 and
HD44179, see Paper~I), which both show PAH emission at the
shorter wavelengths. There is a known PAH feature at about
16.4~$\mu$m (Van Kerckhoven et al. 2000), so we tentatively
identify this feature with PAHs.

\subsection{23 micron complex (22 -- 25.5 $\mu$m)}

\label{sec:ident23}
\begin{figure}[t]
\centerline{\psfig{figure=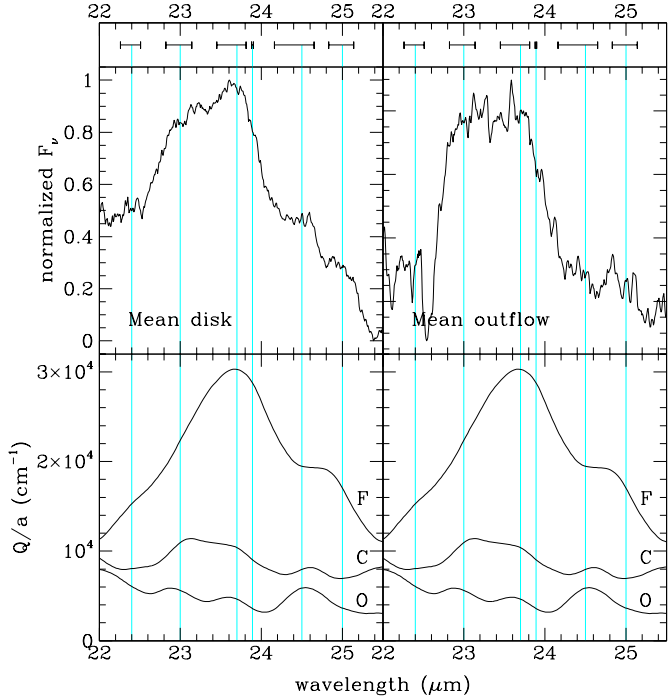,width=88mm,angle=0}}
\caption[]{The normalised mean 23 micron complex spectra together with the
forsterite (F; +1.1E4) and clino-enstatite (C; +7E3) and ortho-enstatite (O) mass absorption coefficient
spectra in this wavelength range. The forsterite and clino-enstatite
spectra are shifted. The gray lines indicate the features found in our
spectra, with their spread in peak position observed in the ISO data indicated above the plot.}
\label{fig:ident23}
\end{figure}

The mean spectra of the 23 micron complex are shown in Fig.~\ref{fig:ident23}
together with the laboratory spectra of forsterite and both types of enstatite.
Clino and ortho-enstatite show similar weak spectral structure, but the
strength ratio of the peaks is reversed.

Both forsterite and enstatite have solid state bands in this
wavelength range. These two components can account for some of the
observed structure. The 23.7~micron feature has been attributed to
forsterite, while the 23.0 and 24.5~micron features coincide with
enstatite features. However, since the strength of these two bands
are not correlated (in the outflow sources the 24.5~micron feature
is weak and the 23.0~micron relatively strong) it is expected
that, at least one, of these two features has a contribution from
another dust species.

The 22.4 and 25.0~micron features still lack identification, they
seem not to be correlated with the other features in this complex.
In most outflow sources we see a drop around 22.5 $\mu$m. Since
absorption at these wavelengths is not expected (for most
sources), it is either caused by emission features at both sides,
or due to responsivity. Since it is not seen in the disk sources
and the two emission features are found in several sources, we
attribute the 22.5~$\mu$m drop to the narrowness of the
crystalline silicate features in the outflow sources.

\subsection{28 micron complex (26.5 -- 31.5 $\mu$m)}
\label{sec:ident28}

\begin{figure}[t]
\centerline{\psfig{figure=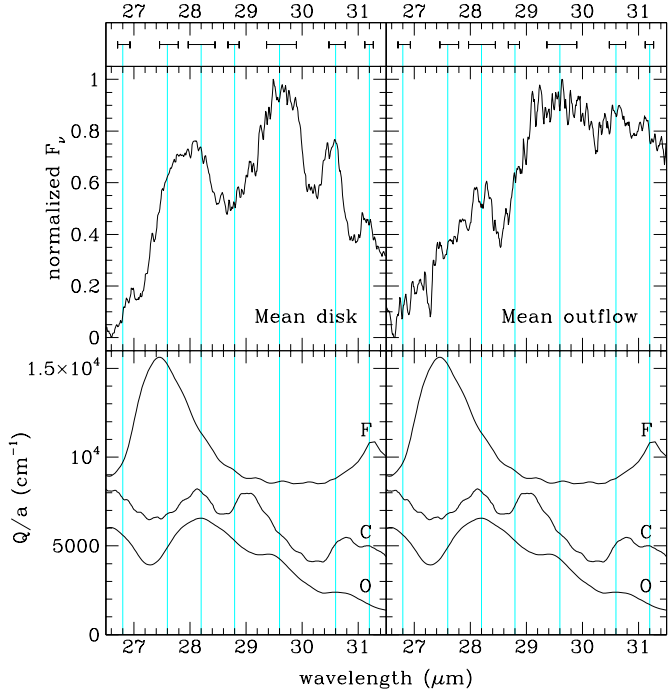,width=88mm,angle=0}}
\caption[]{The normalised mean 28 micron complex spectra together with the
forsterite (F; +8.5E3) and clino-enstatite (C; +3.5E3) and ortho-enstatite (O) mass absorption coefficient
spectra in this wavelength range. The forsterite and clino-enstatite
spectra are shifted. The gray lines indicate the features found in our
spectra, with their spread in peak position observed in the ISO data indicated above the plot.}
\label{fig:ident28}
\end{figure}

The mean spectra of the 28 micron complex are shown in Fig.~\ref{fig:ident28}
together with the laboratory spectra of forsterite and both types of enstatite.
In this complex the spectra of ortho and clino-enstatite start to deviate
slightly from each other.

Laboratory measurements show that forsterite has two features in
this wavelength range: a strong band at 27.5~$\mu$m and a weaker
($\approx 1/3$ of the 27.5~$\mu$m) feature at 31.2~$\mu$m. This
second feature seems to disappear if Fe is included in the lattice
structure (see J\"ager et al. 1998; Koike et al. 1993). The
spectrum of NGC6537 (see Paper I) is remarkable in this respect,
since it does show the typical forsterite features and even a
69.0~micron feature (indicating that the forsterite is very
magnesium rich, (Sec~\ref{sec:60mu} and J\"{a}ger et al. 1998),
while there is no indication of a 31.2~micron feature. It is even
more remarkable considering the fact that the spectrum of NGC6537
is very similar to that of NGC6302 (Paper I) which shows the most
prominent 31.2~micron feature in our sample. This raises some
doubt about the forsterite identification of the 31.2~micron
feature in the ISO spectra and we leave this to future research.

The 28.2~micron feature, often blended with the 27.5~micron
feature, is most likely due to enstatite, although it is not clear
why there is no evidence for an almost equally strong 26.5~micron
feature, which is seen in the laboratory transmission spectra
(J\"ager et al. 1998; Koike et al. 1999). The weak 26.9 and
28.7~micron features as well as the relatively strong 29.6 and
30.6~micron features are not well described by forsterite and/or
enstatite. For the latter two features this is very clear in the
fit to the spectra of NGC6537 and NGC6302 (Paper III). The
strength of 29.6 and 30.6~micron features are correlated. This
would imply that they are originating from the same dust species.
No identification has been made yet.

\subsection{33 micron complex (31.5 -- 37.5 $\mu$m)}
\label{sec:ident33}

\begin{figure}[t]
\centerline{\psfig{figure=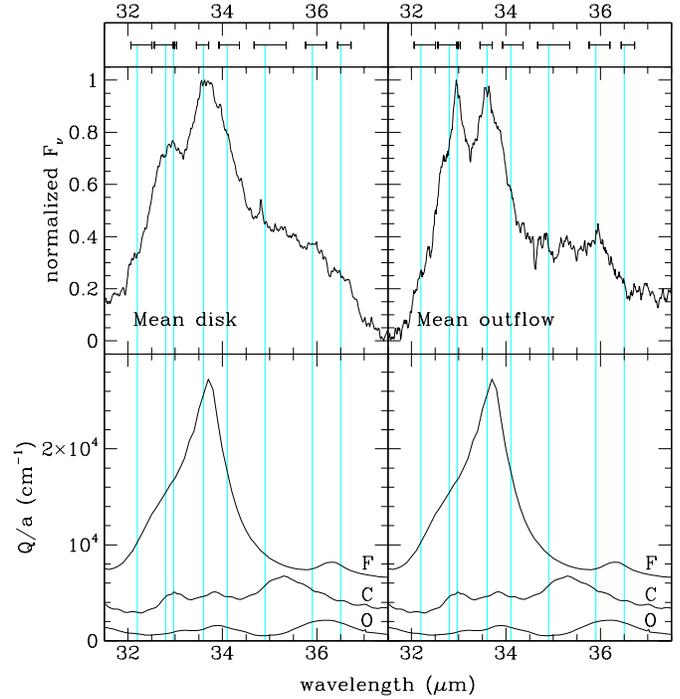,width=88mm,angle=0}}
\caption[]{The normalised mean 33 micron complex spectra together with the
forsterite (F; +6E3) and clino-enstatite (C; +3E3) and ortho-enstatite (O) mass absorption coefficient
spectra in this wavelength range. The forsterite and clino-enstatite
spectra are shifted. The gray lines indicate the features found in our
spectra, with their spread in peak position observed in the ISO data indicated above the plot.}
\label{fig:ident33}
\end{figure}

The mean spectra of the 33~micron complex are shown in
Fig.~\ref{fig:ident33} together with the laboratory spectra of
forsterite and both types of enstatite. There is a small shift
between the clino and ortho-enstatite features, which is most
clear for their peak around 36~$\mu$m.

Both forsterite and enstatite have features in this wavelength
range. The band at 33.6~$\mu$m is identified with the strong
resonance peak of forsterite. The strongest enstatite peaks are
found around 36~$\mu$m (35.3~$\mu$m for clino-enstatite and
36.1~$\mu$m for ortho-enstatite). They might be the relatively
warm carriers of the cold circumstellar 34.9 and 35.9~micron
features. They are likely the main contributors to the plateau
(see also the fits in paper~III). The 36.5~micron peak also falls
into the plateau regime. Forsterite does show a peak at these
wavelengths, which would suggest that the long wavelength end of
the plateau should correlate with the strength of the
33.6~micron forsterite peak. However, this correlation is not
found in our sample. Also, the strength of the laboratory
36.5~micron feature is too weak to explain the strength of the
36.5~micron feature seen in some sources, e.g. MWC922 (Paper~I).
The fits to the plateau in Paper III make clear that we are also
still missing a component that peaks around 34.5~$\mu$m. However
we were not able to extract that feature from our spectra, which
indicates that is is probably very blended with other features.

If the red-edge of the plateau is due to ortho-enstatite, then the
difference in the extension of the plateau (36.5~$\mu$m for the
outflow objects and 37~$\mu$m for the disk sources) could be
explained by a larger percentage of clino versus ortho enstatite.
However, this is in contrast with what has been found in the
analysis of the 40~micron complex (see Sec~\ref{sec:40mu}). So, we
cannot be conclusive about this.

Laboratory spectra of enstatite do show peaks at 33.0 and
34.0~$\mu$m, but these are much weaker than the enstatite features
seen in the plateau. This is probably not a big problem for the
34.1~micron feature. Although the measured 34.1~micron feature is
narrower than seen in the laboratory, this is the case for
more features and might be a temperature effect. The 34.1~micron
feature is a typical example of a feature that often does not show
a sharp peak but seems necessary to add to fit the spectrum. Since
there is no correlation found with the 33.6 or 34.9~micron feature
it is not expected to be related to the non-Gaussian shape of
these features. We note that the 32.97~micron feature in high flux
sources is due to incorrect responsivity curves (see Section~\ref{sec:33mu}).
. The 32.2, and
32.8~micron features still lack solid identification.

\begin{figure}[t]
\centerline{\psfig{figure=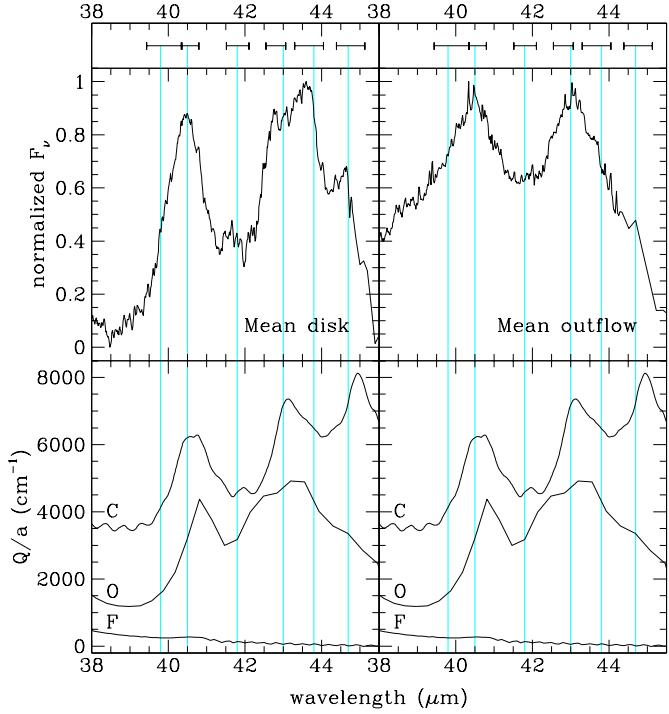,width=88mm,angle=0}}
\caption[]{The normalised mean 40 micron complex spectra together
with the forsterite (F) and clino-enstatite (C; +3.5E3) and
ortho-enstatite (O; +1.2E3) mass absorption coefficient spectra in
this wavelength range. The ortho- and clino-enstatite spectra are
shifted. The gray lines indicate the features found in our
spectra, with their spread in peak position observed in the ISO data indicated above the
plot.} \label{fig:ident40}
\end{figure}

\subsection{40 micron complex (38 -- 45.5 $\mu$m)}
\label{sec:ident40}

In Fig.~\ref{fig:ident40} we show the mean spectra and the
laboratory spectra of forsterite and both types of enstatite. 
Forsterite shows no features in this wavelength region. The
clino-enstatite feature at 44.9~$\mu$m, which we identify with the
44.7~micron feature, is the first feature which is completely
different from the ortho-enstatite features and therefore a good
benchmark for the clino- to ortho-enstatite ratio.

The 40.5 micron feature is probably caused by enstatite, since
laboratory spectra of both ortho- and clino-enstatite show a peak
at this position. The 40.5~micron feature shows large source to
source variations in peak position for such a well defined
feature, the extremes being IRAS09425-6040 (40.35~$\mu$m) and
HD179821 (40.80~$\mu$m) (see Paper~I). This may indicate that two
different dust components are responsible for the emission in the
40.5~$\mu$m range. It should be noted that these two components
are not seen separately, which indicates that, if present, they
severely blend. Although it cannot be excluded at the moment that
it is just one solid state, which can shift quite dramatically, we
have other indications for a second component in the 40.5~micron
feature. The fit to the spectra of HD179821 and AFGL4106 in Paper
III demonstrate that in these cases another dust component is
necessary to explain the strength of the features.

The 43.8 micron feature can probably be explained with
ortho-enstatite. The strength ratio between the 40.5 and
44.7~micron features might therefore give an indication of the
clino-enstatite over ortho-enstatite ratio. However, one would
expect that this difference is also seen at the long wavelength
edge of the 35~micron plateau and there is no correlation found in
that spectral regime. This indicates that other dust species
influence the features attributed to enstatite features, as was
already indicated for the 40.5~micron feature.

The 43.0~micron feature is attributed to crystalline H$_2$O-ice
and clino-enstatite. It should be noted that the wavelength of the
crystalline H$_2$O-ice feature can shift significantly with
temperature, from 42.7 at 10~K to 44.7 at 140~K (Smith et al,
1994). Our peak positions indicate that the average ice
temperature is around 40~K. This is a rather low temperature,
although similar temperatures have been found for the H$_2$O-ice
in AFGL4106 (Molster et al 1999b) and HD161796 (Hoogzaad et al. in prep.),
based on a comparison of the strength and position of the 43 and
60~micron bands of H$_2$O-ice. 
The relative strength of the 43.0~micron feature
in the outflow sources compared to the disk sources suggests that
crystalline H$_2$O-ice is more abundant in the outflow sources
than in the disk sources.

The 39.7~micron feature lacks identification, but there are
indications that its strength shows an anti-correlation with the
strength of the 44.7~micron feature. We therefore speculate that
this feature is due to a reaction product of clino-enstatite and
another (dust) species.

\subsection{60 micron complex (50 -- 72 $\mu$m)}
\label{sec:ident60}

\begin{figure}[t]
\centerline{\psfig{figure=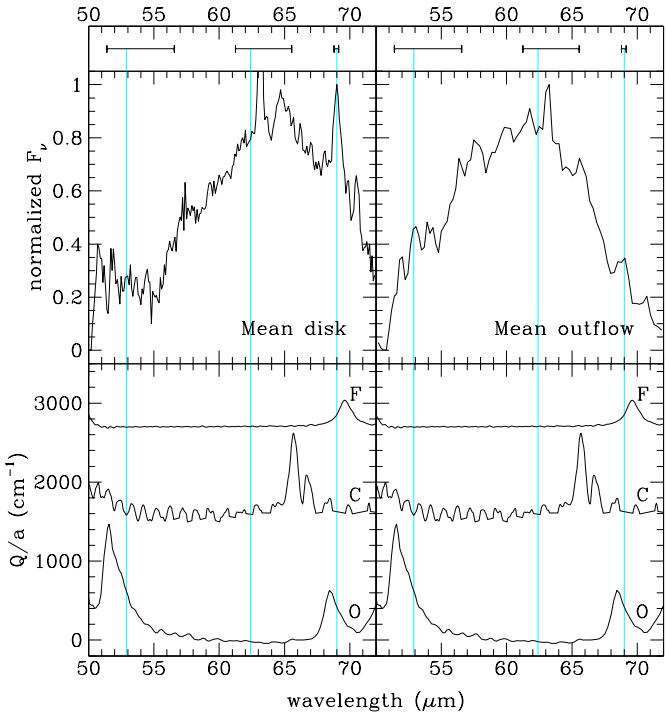,width=88mm,angle=0}}
\caption[]{The normalised mean 60 micron complex spectra together with the
forsterite (F; +2.7E3) and clino-enstatite (C; +1.6E3) and ortho-enstatite (O) mass absorption coefficient
spectra in this wavelength range. The forsterite and clino-enstatite
spectra are shifted. The gray lines indicate the features found in our
spectra, with their spread in peak position observed in the ISO data indicated above the plot.}
\label{fig:ident60}
\end{figure}

Fig.~\ref{fig:ident60} shows the mean spectra of the 60~micron
complex and the corresponding spectra of forsterite and enstatite.
We note that these silicates only show sharp features in their
lab spectra, while this complex is dominated by a broad feature.

The 62~micron feature and the shoulder at 52~$\mu$m, which is
hardly seen in the disk sources and much more prominent in the
outflow sources, is identified with crystalline H$_2$O-ice. In the
disk sources the 62~micron band is red-shifted compared to the
outflow sources. Laboratory data of crystalline H$_2$O-ice show
only a small shift in peak position of the 60~micron band with
temperature. This shift is too small to explain the large
differences in peak position between the disk and the outflow
sources. This indicates that the 62~micron feature actually
contains at least two solid state features: a 60~micron feature
attributed to crystalline water ice and a 65~micron feature. 
The shift to shorter wavelengths in the outflow sources implies that
the relative abundance of crystalline water ice is higher in the
outflow sources, a conclusion which was also drawn on the basis of
the 40~micron complex.
The 65 micron feature is well fitted with diopside 
(Koike et al. 2000).
Enstatite also shows some peaks in the
60-70~$\mu$m wavelength range, but since they are quite sharp in
the laboratory data (Koike et al. 1999) it is not likely that enstatite
is the main contributor to the 65~micron band. 

The 69.0 micron feature has been identified with forsterite. This
peak shifts significantly by adding fayalite to the solid solution
series (already to 73~$\mu$m for a composition of 4\% fayalite and
96\% forsterite, see J\"{a}ger et al. 1998). This feature
therefore indicates that the olivine dust particles are very pure
forsterite with hardly any Fe. However, it should be noted that
the laboratory peak position of pure forsterite at room
temperature is 69.7 $\mu$m, while we observe this feature around
69.0~$\mu$m. Recent laboratory measurements at low temperatures
(Bowey et al. 2000,2001; Chihara et al. 2001) indicate a shift to 
shorter wavelengths with
decreasing temperature; e.g. at 4~K this feature narrows
significantly and peaks at 68.8~$\mu$m. This would provide a
natural solution to the discrepancy between observed and
laboratory data taken at room temperature, see also Paper III.

\subsection{Remaining features}
\label{sec:identrem}

\begin{figure}[t]
\centerline{\psfig{figure=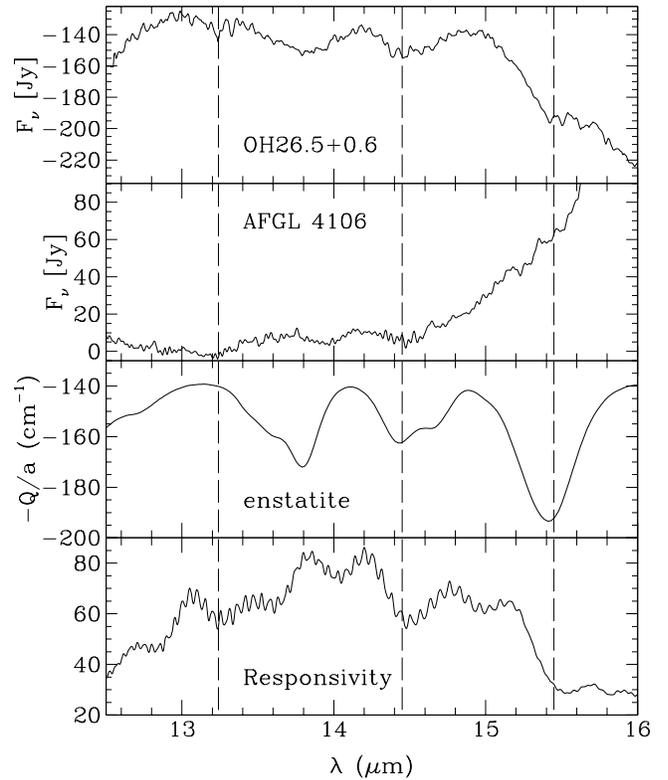,width=85mm}}
\caption[]{A comparison of the features
around 14 $\mu$m in the continuum subtracted
spectra of OH26.5+0.6 and AFGL4106, together with the responsivity profile
and the mirrored (in the wavelength-axis) mass absorption coefficient
spectrum of enstatite.
Three dips in the responsivity are indicated by dashed lines.}
\label{fig:prob14}
\end{figure}

It was noted that the features around 14 $\mu$m might be (partially)
instrumental. The best indication for this statement would seem OH26.5+0.6 which
shows these spurious features also in emission while all other features in this
wavelength range are in absorption. In Fig.~\ref{fig:prob14} we show its
spectrum in this wavelength region, together with the spectrum of AFGL4106.

We would like to point out that the spectrum of OH26.5+0.6 could
be very well fitted by the `absorption' spectrum of enstatite.
However the features around 14 $\mu$m were also found in stars
which only show emission like AFGL4106. In these cases the
enstatite spectrum alone can not explain the structure anymore,
and it looks more like a responsivity problem, also because it
shows quite some similarities with the responsivity spectrum.
Caution should therefore be taken for these features, since the
are likely influenced by both enstatite and responsivity problems.

The 20.7 micron feature is seen in the forsterite and enstatite
data, but not as strong as it is seen in some sources. Silica
(SiO$_2$) does show a very strong resonance at this wavelength and
might therefore be a good candidate. If silica is present it can
be an explanation for the 9.1, 15.9 and 26.1~micron features
(Fabian et al. 2000). However, one would also expect to see
a feature at 12.6 micron which is not the case. Therefore, a solid
identification for this feature is still lacking.

The 21.5 micron feature is also seen as a weak feature in the
enstatite and forsterite laboratory spectra. In some sources it is
stronger than expected from these laboratory spectra. Therefore, it
is likely that another material contributes to this
feature. While the 20.7~micron feature is always present when the
21.5~micron feature is detected, this is not true the other way
around. Therefore, they must originate from different dust
species. Also, the strength ratio is not always the same.

Forsterite has a feature at 26.1~$\mu$m (J\"{a}ger et al. 1998).
We, therefore, identify the 26.1~micron feature with forsterite,
although a contribution from silica cannot be excluded.

The blend of the 47.7 and 48.6~micron features has not yet been
identified; enstatite nor forsterite have features at these
wavelengths. Ferrarotti et al. (2000) identified the blend of the
47.7 and 48.6~micron band with FeSi. The 44.7~micron features
might also be explained by this material. However, their fit is not
completely satisfactory and the FeSi features around 30~$\mu$m
are also not well matched with features in our ISO spectra. We
wonder if shape effects shifts (and reshapes) the bands as is seen
for forsterite. This has not been tested. This identification was
based on a condensation sequence in stellar outflows where free
carbon and oxygen atomes are both underabundant. This can be
either because the C/O ratio is close to~1 and all carbon and
oxygen is locked up into CO, or in stars which expose their
layers, which formerly have burnt hydrogen by the CNO-cycle.
However, we observe the 48~micron feature in several stars, some
of which are definitely O-rich (e.g. OH26.5+0.6). The proposed
identification of the 48 micron blend with FeSi is an interesting
identification, but not convincingly confirmed yet. Other
(hydrous) silicates often have peaks around 48~$\mu$m. However, no
suitable candidate has yet been identified. Therefore, the
identification of the 47.7 and 48.6~micron features remain open.

The longest wavelength feature which has been found in various
stars is the 91~micron feature. 
Malfait et al (1998) suggested hydrous silicates; these materials
often show a broad feature in this wavelength range, but a
satisfying fit has not yet been achieved. It is also difficult to
explain how hydrous silicates would form in these outflows, since
liquid water seems necessary.

In 89~Her and HD44179 a broad (full width zero intensity
$\approx~15~\mu$m) solid state band which peaks around 33~$\mu$m,
hereafter called the 33~micron band, is found. It has not been
included in Table~\ref{tab:complex}, because its shape and appearance
are very uncertain. It shows a relatively sharp rise on the blue
side at about 27~$\mu$m while it ends with a gentle slope on the
red side at about 42~$\mu$m. In the amorphous silicate data
presented by Ossenkopf et al. (1992) a broad feature seems present
around these wavelengths, but the limited wavelength resolution of
this data set at these wavelengths prevents a firm identification.
The strength of this feature in our ISO data is likely too strong
to be solely explained by the feature in the Ossenkopf data set.
Therefore, other materials are likely to contribute to this
feature. The peak position and width bear some similarities with
the 30~micron feature in C-rich sources, where it is attributed to
MgS. Similar features are also seen in the direction of the
galactic center and in the dust of $\eta$~Car. Note that the shape
of the 28 micron complex seems influenced by this feature.

\section{Discussion and conclusions}

In this paper we have combined the spectra of 17~stars with
oxygen-rich circumstellar dust environments. These combined
spectra show a wealth of dust features, which now are studied
in a systematic way. Most of the features cluster into one of
seven complexes (at 10, 18, 23, 28, 33, 40 or 60~$\mu$m). Each of
these complexes is a convolution of a limited set of components,
which vary independently.

Our conclusion from this work is that there is a clear distinction
between the disk and outflow sources. Not only in the strength  of the 
features and therefore in abundance of the crystalline silicates,
which was the original natural separator profile between these two groups
of objects, but also in the characteristics of the complexes.\\
\\
\noindent{\em 10 + 18 micron complex:}\\
The difference in shape, a smooth profile for the outflow sources and 
narrow features for the disk sources, is mainly due to an abundance 
difference between the crystalline and amorphous silicates in the disk and 
outflow sources and is also directly connected to the strength of the features.
The outflow sources have a lower abundance of crystalline silicates.\\
\\
\noindent{\em 23 micron complex:}\\
The strength of the 23.0 micron feature with respect to the strength of the
23.7 micron feature is higher in the outflow sources. The 24.5 micron feature
is more prominent in the disk sources. In Paper III we will demonstrate
that in the (massive) outflow sources, the ratio of enstatite over forsterite 
is higher. This would imply that the 23.0 micron feature results indeed from 
enstatite and that although enstatite has a (weak) feature around 
24.5~$\mu$m, another material is (mainly) responsible for the 
24.5 micron feature.\\
\\
\noindent{\em 28 micron complex:}\\
The 28 micron complex differs significantly from source to source and
several outflow sources are affected by a lack of data. Therefore,
no clear difference could be pointed out. \\
\\
\noindent{\em 33 micron complex:}\\
The features appear sharper in the outflow sources than in the disk sources.
This might have to do with the formation history of the crystalline
silicates (see Paper III).
The (post-)Red Supergiants separate out by almost equally strong 32.8 and
33.6~micron features, even after subtracting the spurious 32.97~micron feature.
This is not a characteristic of all outflow sources.\\
\\
\noindent{\em 40 micron complex:}\\
For the 40 micron complex the main difference is found in the strength ratio
of the 43.0 and 43.8~micron features. In the disk sources, the 43.8~micron
feature is more prominent and in the outflow sources the 43.0~micron feature.
This has to do with the high abundance of crystalline H$_2$O-ice in the outflow
sources.\\
\\
\noindent{\em 60 micron complex:}\\
The main difference in this complex is found in the peak position of the
62~micron feature. This feature is probably a blend of
crystalline H$_2$O-ice, peaking around 60~$\mu$m (the outflow sources),
and diopside plus possibly other crystalline silicates,
peaking around 65~$\mu$m (the disk sources).
Also, the 69.0~micron feature is much stronger in the disk sources than in
the outflow sources, which is probably an abundance effect.\\
\\
It is likely that the different history of these dust grains are
responsible for the spectral differences. MYW suggested that low 
temperature crystallization has taken place in the disk sources. In contrast,
the crystallization of silicates in outflow sources is likely to
have taken place close to the star; i.e., high temperature
crystallization. 

Most features can be explained with forsterite and enstatite. It is clear that
several other features lack a proper identification, and also the strength
ratios are not always correct, suggesting the presence of more contributing
materials. Diopside has been mentioned but the strength of several features
attributed to diopside suggests that it is only a minor component.

The growth of enstatite crystals can occur in a preferential direction
of one of the crystallographic axes. In chondritic porous interplanetary dust
Particles (IDPs), which are believed to be the most pristine material of our
solar system, enstatite has been found in which one or 2 directions of
the crystallographic axes were severely depressed (Bradley et al, 1983).
These enstatite crystals are likely formed from the vapour phase
in a low density hot gas environment (Bradley et al, 1983 and references
therein), in conditions which are similar to the conditions in the
outflows of stars. It is therefore possible that one or two of the
crystallographic axes ([010] and [001] according to what has been seen in
IDPs) are depressed. This will change the emission spectrum of this material
in such a way that certain feature will be weak, corresponding to the
depressed crystallographic axis. This scenario might help to understand why
the relative strength of some features differ from the lab spectra.

Another abundant species is H$_2$O ice. Kouchi and Kuroda (1990) demonstrate
that below 70~K the ice rapidly turns amorphous under the bombardment of UV-
photons. Calculations based on their results and reasonable assumptions for the
UV flux of the central stars would imply that all the crystalline water ice
turns amorphous within a couple of days. Apparently the ice is efficiently
shielded from the UV radiation of these stars. Also the low temperature of the
ice is an indication for the shielding of the crystalline water-ice.
If the material is very clumpy as is seen in e.g. the Helix Nebula 
(first seen by Baade and reported by Verontsov-Velyaminov (1968), it might 
provide a suitable environment to survive the UV radiation.

In the outflow sources the abundance of crystalline water ice seems
much higher than in the disk sources. Assuming that initially the same amount
of H$_2$O was formed, this difference points to a scenario in which 
crystalline water ice is destroyed on time scales longer than the outflow 
time scales. In the outflow of stars, density enhancements
has been found, where molecules and dust are protected from the
harsh UV radiation. This might explain the crystalline water-ice content even
for the cold dust. Also inside disks particles are protected against the UV
radiation, but the turbulence in disks likely causes that most dust particles
will ultimately be exposed to the destructive UV radiation, long before the 
disk has been blown away.

With about 80\% of the features identified, see Table~\ref{tab:complex} 
for the identifications, 
the age of astromineralogy has now really started. Still, about
20\% of the features lack a proper identification. There is
also ample evidence that for a significant amount of the
`identified' features more dust components are necessary to
explain the relative strength of the infrared bands. New laboratory
measurements of cosmic relevant dust species up to at least
100~$\mu$m are required to identify these features. We can now
start to fully exploit the historical information of the
conditions the dust experienced, by studying the composition and
chemical structure of the grains.
The differences found between the different laboratory datasets, are
a significant obstacle in the quantitative interpretation of the spectra. 
The properties of the sample (e.g. the stoichiometry, size of the particles, 
number and size of the individual crystals in a particle) 
should all be well checked before and after the measurents of the optical 
constants. We would also recomend that the different astrophysical 
laboratories exchange their samples in order to have a better understanding
of the possible sources of differences in their datasets.

\vspace{1.0cm}
\noindent{\bf Acknowledgements.}\\
FJM wants to acknowledge the support from NWO under grant 781-71-052 and under
the Talent fellowship programm. LBFMW acknowledges financial support from an 
NWO `Pionier' grant.

\end{document}